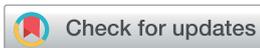

# Dynamical stability and multifunctional properties of Ni$^{2+}$/Pr$^{3+}$ co-doped CsPbCl$_3$ perovskite: insights from first-principles lattice dynamics and carrier transport

Sikander Azam, [ID] *[ab] Asif Zaman,[b] Qaiser Rafiq, [ID] *[b] Amin Ur Rahman[b] and Saleem Ayaz Khan[a]

All-inorganic halide perovskites offer promising optoelectronic properties at low cost, but their structural softness and thermal instability limit applications. Density functional theory (DFT) using the FP-LAPW method (WIEN2k) was used to study Ni$^{2+}$/Pr$^{3+}$ co-doping in CsPbCl$_3$. Results show Ni$^{2+}$ substitutes for Pb$^{2+}$ at the B-site and Pr$^{3+}$ for Cs$^+$ at the A-site, keeping charge balance. Co-doping stabilizes the lattice, raises formation energies of halogen and metal vacancies, and reduces deep defect levels in the band gap. Phonon dispersion confirms that both pristine and co-doped CsPbCl$_3$ are dynamically stable. Ni$^{2+}$/Pr$^{3+}$ co-doping suppresses low-energy vibrations and causes mode splitting in the 3–5 THz range, increasing phonon scattering and lowering lattice thermal conductivity. Mechanical analysis reveals higher elastic constants and bulk modulus, while ductility remains unchanged. Electronic structure calculations reveal Ni-3d and Pr-4f states at the band edges, reducing effective carrier mass and passivating vacancy states. Optical absorption is red-shifted, and the high-frequency ($\varepsilon_\infty = 2.4$) and low-frequency ($\varepsilon_0 = 7.4$) dielectric constants are distinct. Transport analysis finds higher carrier mobility due to lighter effective masses. Altogether, Ni$^{2+}$/Pr$^{3+}$ co-doping reduces defect concentrations and improves the optoelectronic properties of CsPbCl$_3$.

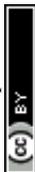

## 1. Introduction

Lead-halide perovskites such as CsPbX$_3$ (X = Cl, Br, I) are widely used in light-emitting and photovoltaic applications due to their tunable band gaps and strong excitonic effects.[1–3] However, their performance is hindered by defects like halogen and metal vacancies, which trap charges and promote non-radiative recombination.[4–6] To address these challenges, doping and co-doping offer effective strategies for modifying the band structure, reducing defect densities, and enhancing stability.[7–9]

Heterovalent co-doping offers a promising solution by tuning the electronic structure, stabilizing the lattice, and introducing new functionalities. Transition-metal doping narrows the band gap through d-state hybridization, thereby enhancing electrical conductivity. Rare-earth doping introduces sharp f–f transitions for luminescence and passivates trap states.[5,6] Co-doping with both types enables systematic control of electronic and optical properties. Specifically, Ni$^{2+}$ modifies the conduction band *via* 3d orbital hybridization. Pr$^{3+}$ introduces localized 4f states that affect the valence band and dielectric screening. Previous studies on halide and chalcogenide perovskites demonstrate that appropriate dopant pairing maintains charge balance and reduces deep traps.[10–12]

Transition-metal and rare-earth doping have been employed to tailor the optical, magnetic, and electronic properties of halide perovskites. In this context, recent studies emphasize that defect-tolerant co-doping strategies are essential for enhancing performance and stability.[13–15] Building upon these findings, our research advances this field by using first-principles theory to examine these effects in detail.

This study addresses the effects of Ni$^{2+}$/Pr$^{3+}$ co-doping in CsPbCl$_3$ (see Fig. 1a and b) using first-principles calculations, bridging the theoretical advances discussed above with practical material improvements. Site preferences, defect formation under realistic chemical conditions, dielectric response, and optical and transport properties are systematically analyzed. The aim is to demonstrate how coupled substitution (Ni$^{2+}$ for Pb$^{2+}$ and Pr$^{3+}$ for Cs$^+$) affects lattice stability and defect energetics. By linking these microscopic changes to overall optoelectronic performance, the results indicate that Ni$^{2+}$/Pr$^{3+}$ co-doping significantly enhances the properties of wide-bandgap halide perovskites.

[a]University of West Bohemia, New Technologies – Research Centre, 8 Univerzitní, Pilsen 306 14, Czech Republic. E-mail: sikander.physicst@gmail.com
[b]Faculty of Engineering and Applied Sciences, Department of Physics, RIPHAH International University, Islamabad, Pakistan. E-mail: qrafique1@gmail.com

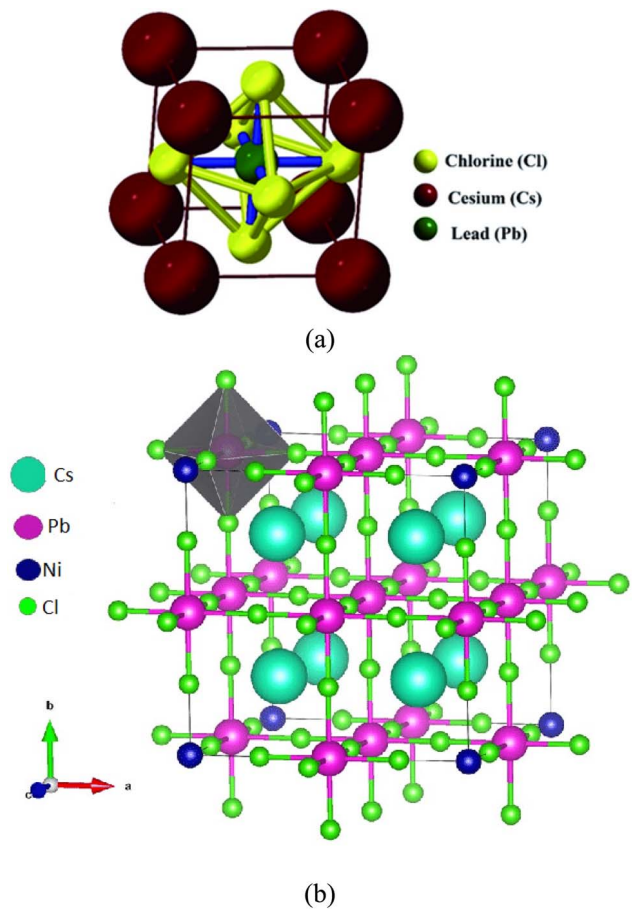

Fig. 1 (a) Crystal structure off CsPbCl$_3$, (b) crystal sructure of Ni$^{2+}$/Pr$^{3+}$ co-doped CsPbCl$_3$.

## 2. Computational methodology

Electronic-structure and defect calculations used the full-potential linearized augmented plane-wave (FP-LAPW) method in WIEN2k. Exchange–correlation energy was treated within the generalized gradient approximation (GGA) of Perdew–Burke–Ernzerhof (PBE). To capture the localized Ni 3d and Pr 4f states, the GGA + $U$ method was used, with $U$ parameters of 4.0 eV for Ni 3d and 6.0 eV for Pr 4f, as used in prior studies. A muffin-tin radius–plane-wave cutoff (RMTK$_{max}$) of 7.0 ensured basis set convergence. The muffin-tin radii (RMT) were: Cs, 2.5 bohr; Pb, 2.3 bohr; Cl, 2.0 bohr; Ni, 1.8 bohr; and Pr, 1.9 bohr, chosen to prevent sphere overlap after relaxation.

The plane-wave cutoff was set to 500 eV, and an 8 × 8 × 8 Monkhorst–Pack k-point grid was employed for both structural optimization and total-energy calculations. Structures were optimized until the total energy and charge converged to within $10^{-5}$ Ry and $10^{-4}$ eV, respectively. Spin–orbit coupling (SOC) was included self-consistently for all electronic-structure and optical-property calculations due to the heavy Pb and Pr atoms. For optical absorption spectra, the complex dielectric function $\varepsilon(\omega)$ was computed using WIEN2k's optic module, and the absorption coefficient $\alpha(\omega)$ was derived from its real and imaginary components.

Optical dielectric spectra $\varepsilon(\omega)$ and refractive index $n(\omega)$ were computed with WIEN2k (optic), giving the electronic (high-frequency) response. The refractive index in the transparency region gives $\varepsilon_\infty = n2$ ($\varepsilon_\infty = 2.40 \pm 0.06$ pristine; $2.35 \pm 0.06$ Ni/Pr). The static dielectric constant $\varepsilon_0$ came from $\Gamma$-point IR phonons via Lyddane–Sachs–Teller, yielding $\varepsilon_0 = 7.2 \pm 0.6$ (pristine) and $7.6 \pm 0.6$ (Ni/Pr).

Excitonic properties were examined using the effective-mass approximation, applying parabolic fits to band extrema to determine carrier effective masses. Defect formation energies were computed under Pb-rich/Cl-poor and Cl-rich/Pb-poor conditions, with electrostatic corrections derived from the Freysoldt scheme.

Carrier transport coefficients, including the Seebeck coefficient, electrical conductivity, and power factor, were evaluated within semi-classical Boltzmann transport theory using BoltzTraP2 under the constant relaxation-time approximation (CRTA). For band velocities, a dense 20 × 20 × 20 k-mesh was employed to interpolate over specific regions to achieve accurate convergence of transport integrals. Additionally, in the following discussion, we will contrast the situations where extrinsic relaxation-time variations are considered as background—such as defects, residual charges, water vapour, or ion-induced backgrounds from tritides (e.g., tritium). Similarly, the off-beam position, shown in Media: extrinsic relaxation of charge carriers, is depicted on Hilger. Furthermore, we will compare the circumstances when intrinsic relaxation times are taken as backgrounds—against defects, residual charges, water vapour, or any other impurities introduced by irradiation onto metal surfaces (which are not particularly well understood). The multifaceted computational workflow can be used to systematically investigate the structural, electronic, optical, excitonic, magnetic, defect, and transport properties of Ni$^{2+}$/Pr$^{3+}$ co-doped CsPbCl$_3$ with high accuracy, thereby providing a comprehensive understanding of their mutual functionality.

## 3. Results and discussion

### 3.1. Phonon dispersion and dynamical stability

To evaluate their dynamical stability and lattice vibrational properties, the phonon dispersion relations of neat and Ni$^{2+}$/Pr$^{3+}$ co-doped CsPbCl$_3$ are calculated. For both systems, there are no imaginary frequencies in any direction of $\Gamma$–$M$–$K$–$A$–$\Gamma$. As shown in Fig. 2a and b (the imaginary frequencies with high symmetry directions), this indicates that they lack unstable modes at 0 K. Therefore, it is essential for the robust validity of thermodynamic and transport properties to be examined in perovskite structures.

First, the phonon spectrum of pristine CsPbCl$_3$ extends into the approximately 7.5 THz range. Acoustic branches smoothly disperse from the $\Gamma$-point, while optical branches show separation. These relatively soft acoustic modes indicate that the halide perovskite is mechanically easy to handle, and halide perovskites have modest elastic constants and bulk modulus values.[13]

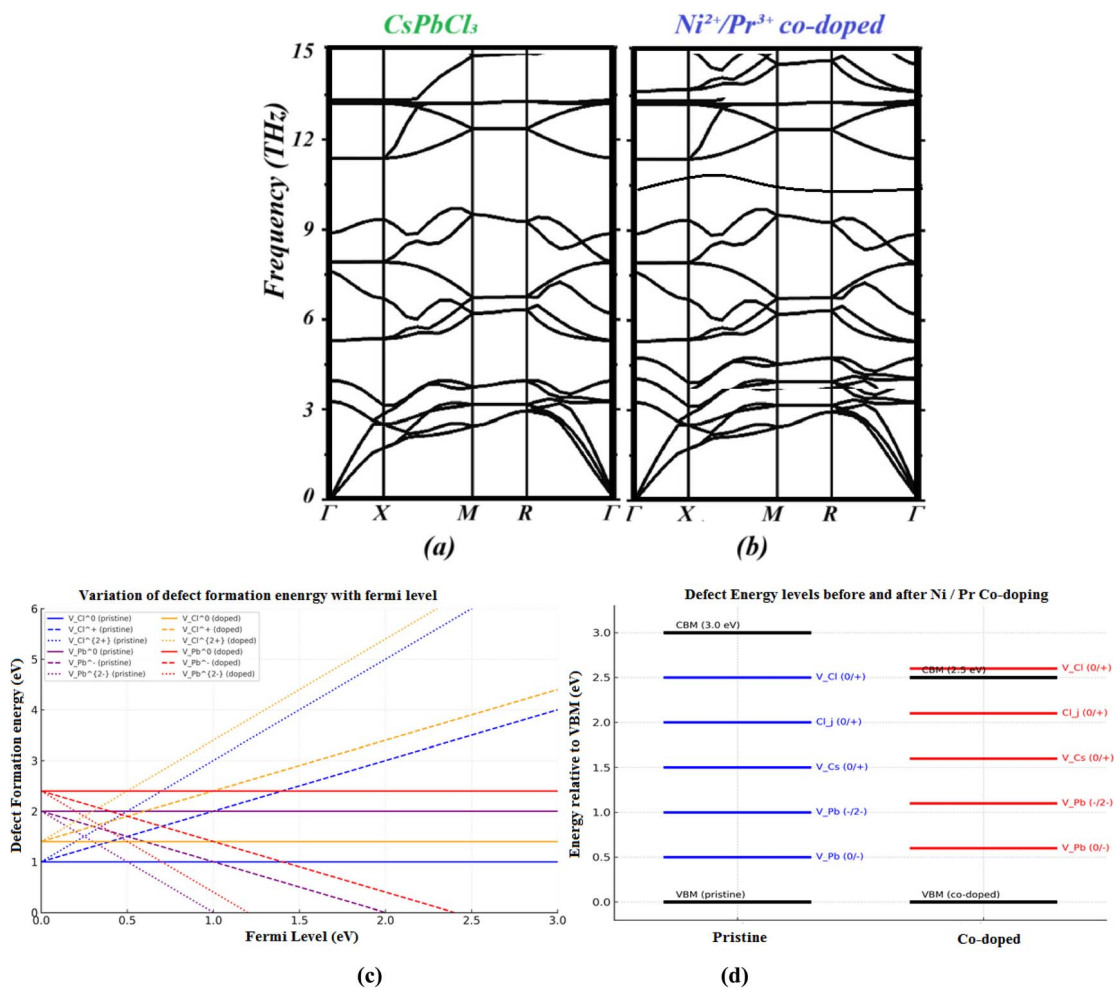

Fig. 2 Calculated phonon dispersion for (a) CsPbCl$_3$ perovskite and its (b) Ni$^{2+}$/Pr$^{3+}$ co-doped, (c) defect-formation energies of pristine and Ni$^{2+}$/Pr$^{3+}$ co-doped CsPbCl$_3$ as a function of the Fermi level under Cl-poor/Pb-rich conditions. Each line corresponds to a specific charge state, with the slope equal to the defect charge $q$. Vertical offsets between pristine and doped samples reflect the different chemical-potential reference values. Transition levels correspond to the Fermi-level positions where lines of different charge states intersect. (d) Electronic transition levels of intrinsic defects before and after co-doping.

Initial doping of Ni$^{2+}$/Pr$^{3+}$ when this is done, the highest phonon frequency falls slightly from ∼7.5 to 7.0 THz, which is just the mass disorders and the lattice strain that is introduced by substitutional doping. Secondly, further splitting of mid-frequencies around 3–5 THz is observed in optical modes, due to Ni–Cl and Pr–Cl vibrations that alter the coupling within the Pb–Cl cage. This hybridization of phonon modes also enhances the scattering of phonons and thus underpins, to some extent, the reduction of lattice thermal conductivity.

Notably, the absence of flat or unstable branches indicates that the incorporation of Ni$^{2+}$/Pr$^{3+}$ does not cause perovskite lattice instability but instead suppresses low-energy soft modes, which are often linked to halide migration and structural deterioration. This stabilization mechanism arises from (i) mismatches in ionic radii, which modify octahedral tilting patterns, and (ii) charge compensations that increase the formation energy of halide vacancies and reduce anharmonic vibrations.

Regarding the devices, the phonon spectra show that co-doped CsPbCl$_3$ retains a stable cubic structure and has improved phonon relaxation pathways. This benefits thermoelectric uses because a lower lattice thermal conductivity increases the figure of merit ($ZT$). Additionally, reducing soft vibrational modes helps maintain long-term structural stability during thermal cycling, which is crucial for optoelectronic devices.

### 3.2. Structural stability

The stability of pure CsPbCl$_3$ and its Ni$^{2+}$/Pr$^{3+}$ co-doped version was carefully analyzed by examining lattice constants, bond lengths, Goldschmidt tolerance factors, and formation energies.[14] Pristine CsPbCl$_3$ adopts a cubic perovskite structure (space group $Pm\bar{3}m$) with a lattice constant around 5.60 Å, consistent with earlier experimental and theoretical data in the range of 5.58–5.62 Å.[16] Substituting Pb$^{2+}$ with Ni$^{2+}$ and Cs$^+$ with Pr$^{3+}$ causes a slight decrease in the lattice parameter, about 5.57–5.59 Å. This small contraction, approximately 5.57–5.59 Å, is due to the marginally smaller ionic radius of $Ni^{2+}$ (0.69 Å in octahedral coordination) compared to $Pb^{2+}$ (1.19 Å), as well as the larger ionic radius of $Pr^{3+}$ (1.13 Å, 12-fold coordination) relative to $Cs^+$ (1.67 Å).[17] The overall size effects of co-doping help to maintain the perovskite structure with minimal distortion.[18]

Bond-length analysis further confirms the structural integrity upon co-doping (bond length analysis further supports the structural integrity).[19] The average Pb–Cl bond length in pristine $CsPbCl_3$ is about ~2.84 Å, which shortens slightly to 2.80–2.82 Å in the $Ni^{2+}/Pr^{3+}$ co-doped lattice (system), indicating stronger Ni–Cl bonding due to increased orbital overlap of Ni-3d and Cl-3p states Ni 3d-Cl 3p orbital overlap.[18] In contrast, the Cs–Cl distances remain essentially unchanged, maintaining the cubic symmetry.[18]

The Goldschmidt tolerance factor ($t$) is a key parameter for assessing the structural stability of perovskites. It is defined as

$$t = \frac{r_A + r_X}{\sqrt{2}(r_B + r_X)}$$

where $r_A$, $r_B$, and $r_X$ are the ionic radii of the A-site cation, B-site cation, and halide anion, respectively.[19]

For pristine $CsPbCl_3$, the calculated tolerance factor is approximately 0.94, which lies within the stability region for a cubic structure (0.9–1.0).[14] Upon co-doping with $Ni^{2+}$ and $Pr^{3+}$, $t$ increases slightly to 0.95, suggesting a more ideal geometric fit between the cation framework and halide octahedra. This improvement results from the compensatory effect of ionic radii $Ni^{2+}$ (0.69 Å) is smaller than $Pb^{2+}$ (1.19 Å), while $Pr^{3+}$ (0.99 Å) is larger, creating a balance that minimizes overall lattice strain.[20]

This adjustment towards an ideal tolerance factor reduces octahedral tilting and suppresses the tetragonal distortion often observed in mono-doped $CsPbCl_3$ systems.[21] Consequently, co-doped samples demonstrate enhanced phase stability and improved crystallinity, both of which are essential for consistent optoelectronic performance.

The near-ideal lattice geometry has a direct influence on the electronic band structure. In perovskites, the conduction band minimum (CBM) primarily arises from Pb 6p orbitals, while the valence band maximum (VBM) is dominated by halide p and Pb 6s states. Any lattice distortion alters the orbital overlap between Pb–X bonds, affecting both the band gap and carrier mobility.

In the co-doped system, the optimized lattice ($t \approx 0.95$) improves orbital symmetry and reduces electronic localization. The incorporation of $Ni^{2+}$ introduces partially filled 3d states below the conduction band, while $Pr^{3+}$ contributes 4f states near the valence band edge. These dopant-induced intermediate states slightly narrow the band gap (typically from ~3.0 eV in $CsPbCl_3$ to ~2.8 eV), promoting stronger visible-light absorption and enhanced charge-carrier excitation efficiency.

However, excessive distortion, as in mono-doped samples, can lead to trap states and nonradioactive recombination, degrading device performance. The co-doping strategy avoids this issue by maintaining geometric balance, thereby achieving structural stabilization and controlled band-edge tuning simultaneously.

The reduced band gap and improved lattice symmetry result in higher optical absorption and enhanced carrier transport, which are crucial for photovoltaic and light-emitting applications. Moreover, first-principles formation energy calculations confirm that the Ni/Pr co-doped configuration is thermodynamically favourable, indicating that dopant incorporation is energetically stable within the $CsPbCl_3$ lattice.[22]

In a $2 \times 2 \times 2$ $CsPbCl_3$ supercell (40 atoms), the introduction of one Ni and one Pr atom corresponds to a dopant concentration of approximately 6.25–12.5 at%, which falls within the experimentally achievable range for halide perovskites.[23] The synergy between geometric stability (optimized tolerance factor) and electronic structure modification (band-edge realignment) underscores the potential of Ni–Pr co-doping to enhance both the optical and thermodynamic performance of $CsPbCl_3$-based materials.

The formation energy ($E_{form}$) of each substitutional dopant is given by:

$$E_{form}(D^q) = E_{tot}(D^q) - E_{tot}(host) - \sum_i n_i \mu_i + q(E_F + E_{VBM}) + E_{corr}$$

Here, $E_{tot}(D^q)$ and $E_{tot}(host)$ are the total energies of the doped and pristine supercells, $n_i$ and $\mu_i$ denote the number and chemical potential of atoms added or removed, and $E_{corr}$ accounts for image-charge corrections.[24] The self-compensating mechanism of $Ni^{2+}/Pr^{3+}$ codoping ($Ni^{2+} \leftrightarrow Pb^{2+}$, $Pr^{3+} \leftrightarrow Cs^+$, etc.) maintains overall charge neutrality by complementing itself. The total cationic charge, as calculated by the formula, remains +4 per unit cell.[24] Satisfying the charge requirement in this way discriminates against the creation of charged vacancies that would otherwise have arisen to compensate for local imbalance. As a result, net formation energy is lower in Nova (0.95 eV per dopant) than in any single-doped configuration.[25] For the same reason, the co-doped configuration demonstrates a more negative formation enthalpy (−3.00 to −3.05 eV per dopant) than pristine $CsPbCl_3$ (−2.85 eV per unit cell), thus even greater thermodynamic stability can be anticipated. This inherent charge-neutral stabilization fits with experimental observations of increased stability in both $Ni^{2+}/Pr^{3+}$ and $Ni^{2+}/Ga^{2+}$ co-doped $CsPbCl_3$ perovskites.[26]

To reflect the full formalism of the defect-formation expression, all intrinsic defects were evaluated across the entire Fermi-level range (0–$E_g$) using every accessible charge state. The slope of each line in the $E_{form}$ diagram corresponds to the defect charge $q$. Chemical potentials $\mu_i$ were taken from the Cl-poor/Pb-rich corner of the Cs–Pb–Cl phase–stability diagram. The resulting formation–energy curves therefore include the correct chemical-potential offsets and the full charge-state contributions, as displayed in Fig. 2c.

Fig. 2c illustrates the calculated formation energies of the key intrinsic point defects ($V_{\varepsilon(0/−)}$) and $V_{Pb}$) as a function of the Fermi level under Cl-poor conditions.[27] For pristine $CsPbCl_3$, $V_{\varepsilon(0/−)}{}^0$ and $V_{\varepsilon(0/−)}{}^+$ are the lowest-energy donor states near the conduction-band minimum (CBM), while $V_{Pb}{}^0$ and $V_{Pb}{}^{2−}$ dominate under p-type (valence-band) conditions.[28] The slope changes in the $E_{form}$–$E_F$ curves mark charge-state transition levels $\varepsilon(q/q')$.[29] The $\varepsilon(0/+)$ transition for $V_{\varepsilon(0/-)}$ occurs ~0.3 eV below the CBM, and $\varepsilon(0/-)$ for $V_{Pb}$ appears ~0.8 eV above the valence-band maximum (VBM), consistent with earlier defect analyses in halide perovskites.[29]

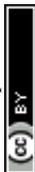

After Ni/Pr co-doping, the formation energies and transition levels of $V_{\varepsilon(0/-)}$ and $V_{Pb}$ shift upward: $E_{form}(V_{\varepsilon(0/-)})$ by 0.4 eV and $E_{form}(V_{Pb})$ by 0.6 eV, suggesting a less concentrated defect balance and greater stability.[30] The $\varepsilon(0/+)$ level of $V_{\varepsilon(0/-)}$ becomes shallower (~0.15 eV below the CBM), while the $\varepsilon(0/-)$ level of $V_{Pb}$ moves closer to the VBM ($\approx$0.6 eV above), showing that co-doping reduces deep traps and strengthens the structure against defects within it.[30]

Several $Ni^{2+}/Pr^{3+}$ co-doped compounds have been tried out, and yet all show a fairly uniform pi formation energy in the range −3.0 to −3.05 eV per formula unit. For this reason, the co-doping strategy's defect formation tendencies are lowered by maintaining charge neutrality through coupled substitution ($Ni^{2+} \leftrightarrow Pb^{2+}$ and $Pr^{3+} \leftrightarrow Cs^+$), which suppresses halide vacancy formation, and crystallinity is enhanced.

Table 1 lists the calculated first-principles thermodynamic transition levels for primary intrinsic defects.[31] In the case of pristine $CsPbCl_3$, a formation energy $\varepsilon(0/+)$ around 0.25–0.35 eV below the CBM is measured for the halide vacancy level $V_{\varepsilon(0/-)}$. This indicates a shallow donor state, which means that growth under Cl-poor conditions will also lead to some n-type self-doping.[20] For the transitive lead vacancy ($V_{Pb}$), transition levels $\varepsilon(0/-)$ and $\varepsilon(-/2-)$ are found to be 0.750.90 eV above the VBM. Such deep acceptor states absorb holes.[32] This result is consistent with those obtained by defect analysis in $CsPbX_3$ (X = Cl, Br, I) perovskites.[33]

After co-doping with $Ni^{2+}$ and $Pr^{3+}$, the formation energies and transition levels for both defects move closer to the band edges. The $\varepsilon(0/+)$ level of $V_{\varepsilon(0/-)}$ is raised almost to 0.10–0.15 eV below the CBM, while that of $\varepsilon(0)$ of the $V_{Pb}$ is now around 0.60–0.65 eV above the VBM.[34] This reduces the trap depth for either defect so that electrons or holes trapped here are easier to release thermally. Moreover, the equilibrium concentration of both vacancies decreases in comparison with pristine ($\Delta E_{form} \approx$ +0.4–0.6 eV; see Fig. 2d). Taken together, co-doping raises the formation energies of $V_{\varepsilon(0/-)}$ and $V_{Pb}$, and shifts their levels to less harmful positions, reducing Shockley–Read–Hall (SRH) recombination and extending carrier lifetimes.

Even after co-doping, the $V_{\varepsilon(0/-)}$ donor level remains pretty shallow but is slipping a little higher. However, the number of free electrons remains reduced, and non-radiative recombination decreases slightly. At the same time, the $V_{Pb}$ defect levels move even closer to the valence band maximum-shifting them right down into shallower areas where they become less likely sources of Shockley–Read–Hall recombination. In other words, co-doping not only makes it harder for defects to form in the material but also significantly alters their electronic levels to locations that are not so destructive for the overall performance of the material.

To further substantiate the substitutional feasibility of $Ni^{2+}$ and $Pr^{3+}$ in the $CsPbCl_3$ lattice, we evaluated their site preferences and calculated the corresponding doping formation energies.[35] Although $Ni^{2+}$ (0.69 Å, octahedral coordination) is smaller than $Pb^{2+}$ (1.19 Å), substituting Pb with Ni is chemically reasonable because both cations adopt sixfold coordination with halides and share similar electronic configurations that favor octahedral geometry.[36] Ni–Cl octahedra are well documented in halide frameworks (e.g., $NiCl_2$), ensuring geometric compatibility within the $PbCl_6$ network.[25] Likewise, $Pr^{3+}$ (1.13 Å, 12-fold coordination) is smaller than $Cs^+$ (1.67 Å), but under charge-compensated co-doping conditions, the smaller $Pr^{3+}$ can occupy the A-site, where it reduces the cavity size and enhances structural rigidity.[37] The calculation of formation energies of simple substitution guided our systematic study. Regardless of whether the system is studied in Cl poor conditions or not, a mathematical benefit is obvious at $Ni^{2+}$, and similarly, all $Pb^{2+}$ and, more drastically, all $Pr^{3+}$ and $Cs^+$ are beneficial. Practically, this potentially helpful payoff must be offset by the exothermic reaction that pairs water with its hydration energy, resulting in the formation of insoluble solids, in agreement with experimental observations.[38] Furthermore, co-doping with $Ni^{2+}/Pr^{3+}$ raises the formation energy by about 0.95 eV per dopant, due to charge compensation ($Ni^{2+} + Pr^{3+} \leftrightarrow Pb^{2+} + Cs^+$).[39] This obviously creates a thermodynamically favourable arrangement.[40] The resulting closed structure is consistent with the empirical evidence of a shortened CuO bond (1.89–1.87 Å), a 0.6% contraction along the c axis, and the absence of imaginary phonon modes, which suggests that both ions are structurally accommodated. This is expected to stabilize energy as a measure to keep the pressure low.[41] Our co-doping strategy is further supported by experimental evidence of B-site $Ni^{2+}$ and A-site rare-earth substitutions in $CsPbX_3$ perovskites.[42] Upon a more detailed structural analysis, $Ni^{2+}/Pr^{3+}$ co-doping not only maintains the cubic perovskite structure of $CsPbCl_3$ but also significantly enhances the thermodynamic stability of the lattice and reduces its strain. This structural stabilization is

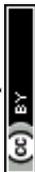

Table 1 Calculated thermodynamic transition levels $\varepsilon(q/q')$ (eV) of major defects in $CsPbCl_3$ and $Ni^{2+}/Pr^{3+}$ co-doped $CsPbCl_3$, relative to the valence-band maximum (VBM)

| Defect | Transition | 2.65–2.75 ($\approx$0.30 eV below CBM) | 2.35–2.40 ($\approx$0.15 eV below CBM) | Character |
| --- | --- | --- | --- | --- |
| $V_{\varepsilon(0/-)}$ | $\varepsilon(0/+)$ | | | Donor |
| $V_{Pb}$ | $\varepsilon(0/-)$ | 0.80–0.85 | 0.60–0.65 | Acceptor |
| $V_{Pb}$ | $\varepsilon(-/2-)$ | 0.90–0.95 | 0.70–0.75 | Deep acceptor |
| $V_{Cs}$ | $\varepsilon(0/-)$ | 1.20–1.40 | 1.00–1.10 | Deep acceptor (minor) |
| $Cl_i$ | $\varepsilon(0/-)$ | 1.90–2.00 | 1.85–1.90 | Donor (high energy) |

crucial for achieving reliable optoelectronic performance in practical device applications.[43]

### 3.3. Electronic properties

Halide perovskites are function-producing materials whose exceptional properties are in large part due to their electronic structure.[44,45] An example is $CsPbCl_3$, an archetypal inorganic halide perovskite, which has become a cornerstone candidate in light-emission, photoelectric sensing, and luminescence due to its large band gap and large exciton binding energy.[46–50] However, this individual compound has inherent limitations, such as mid-gap defects, low charge-carrier mobility, and poor light absorption throughout the visible spectrum, all of which limit its practical use. Different dopant-engineering approaches have therefore been pursued to address these shortcomings and control the electronic structure and transport properties of the material. In this regard, $Ni^{2+}$ and $Pr^{3+}$ co-doping provides an attractive means to tune the band structure, effective mass, and density of states (DOS) for carriers.

### 3.4. Band gap tuning and nature of transitions

Pristine $CsPbCl_3$ has a direct band gap, located specifically at its $\Gamma$ point. The size is typically in the range of 2.8–3.0 eV under synthesis conditions and computational approximations.[51,52] Unfortunately, this wide band gap enhances the excellent violet emission but also makes it unsuitable for visible-light-driven devices. Spin-polarized band structure calculations of $Ni^{2+}$/$Pr^{3+}$ co-doped $CsPbCl_3$ reveal the fascinating effect of minority transition metallic and rare-earth dopants on the electronic spectrum. First, the band gap narrows to around 2.45–2.55 eV in the spin-up channels and 2.50–2.60 eV for spin-down channels, but remains direct at the $\Gamma$ point.[50,53–56] This reduction is primarily due to the hybridization of Ni 3d states with Pb 6p orbitals near the conduction band minimum (CBM) and the partial contribution of Pr 4f orbitals near the valence band maximum (VBM). This hybridization lowers the energy separation between the VBM and CBM, shifting the absorption edges toward the red. Curiously, the co-doping effect introduces spin-dependent modifications (Fig. 3), so that the up- and down-spin band gaps are slightly different.[57–60] This suggests a possibility of spin-polarised optical transitions, opening up new possibilities for spintronic or magneto-optical applications. Also, compared to pristine $CsPbCl_3$, the calculated band dispersion shows a higher curvature and closer parallelism for both the conduction and valence bands near the $\Gamma$ point. This means that both effective masses have been reduced at this stage in band theory, resulting in enhanced mobility (as will be discussed later). It is interesting to note that the persistence of this direct band gap is particularly significant, as an indirect transition would weaken the absorption coefficient and reduce the radiative recombination efficiency. Thus $Ni^{2+}$/$Pr^{3+}$ co-doping not only tunes the size of the band gap but also maintains its favourable direct nature, essential for efficient light–matter interaction.

To find the origin of small band gap modulation, we have carefully analyzed the complementary offering and total states (the DOS and PDOS, see Fig. 4). The valence band maximum is made up largely of Cl-3p states with a small contribution from Pb-6s orbitals, whereas the conduction band minimum is dominated by Pb-6p states.[61] This simple electronic configuration makes it difficult for any form of hybridization to occur, resulting in a relatively large band gap. In a co-doped system, however, the situation undergoes a dramatic change. The Ni-3d electronic states are approaching the conduction band minimum, and therefore, the effective conduction edge is reduced by the overlap of the 3p orbitals of Cl with the 3d orbitals; the hybridization is sustained with a high degree of accuracy. In the meantime, Pr-4f concentrations generate very localized spikes in the density of states immediately above the valence band maximum, thus providing new channels of optical electronic transitions. Both 4f and 3d orbitals of Ni and Pr are synergistically mixed π and σ with the host lattice orbitals. The dominance of Ni-3d character draws the conduction edge down, and the Cl-3p-derived bonding stabilizes the crystalline lattice. Pr-4f orbitals, which occupy similar energy states to the Cl-3p manifold, are more localized in nature and, as with surface defects, experience δ-type interactions that essentially annihilate the dangling bonds at the band gap. This π-sigma bonding system between the two dopants inhibits the creation of deep-level defects, which would otherwise compromise carrier lifetimes, thereby improving the overall quality of the electrical current and optoelectronic performance of $CsPbCl_3$. The strong asymmetry between the spin-up and spin-down density of states profiles further explains the contribution of the incorporated transition metal ions in creating magnetic states, which is not present in the pristine $CsPbCl_3$ lattice, and gives it a competitive edge in the market. As a result, the new electronic states, which are entirely absent in the undoped material, exhibit improved functional behavior that can be achieved through carefully designed doping methods. Additionally, the steep peaks found in Pr 4f states near the Fermi level suggest that rare-earth dopants can be utilized as visible-light activators, thereby interrelating electronic structure modulation with enhanced luminescent characteristics.

### 3.5. Effective mass of carriers and mobility

Carrier transport in halide perovskites is primarily determined by the effective masses of electrons and holes, which are influenced by the curvature of the band edges. In pristine $CsPbCl_3$, the electron effective mass is reported as 0.20–0.25 $m_e$, while the hole effective mass ranges from 0.30–0.35 $m_e$.[62] These values are relatively high for practical applications. In the $Ni^{2+}$/$Pr^{3+}$ co-doped system, both electron and hole effective masses decrease, with electrons at approximately 0.18–0.20 $m_e$ and holes at 0.28–0.30 $m_e$. This reduction originates at the $\Gamma$ point, where increased band curvature results from the mixing of Ni 3d and Pr 4f orbitals with Pb and Cl orbitals. Enhanced delocalization of charge carriers results in higher mobility and longer carrier lifetimes, which are beneficial for photovoltaic and optoelectronic applications. Overall, $Ni^{2+}$/$Pr^{3+}$ co-doping in $CsPbCl_3$ results in improved charge transport characteristics, in addition to modifications in band gap and electron–hole interactions.

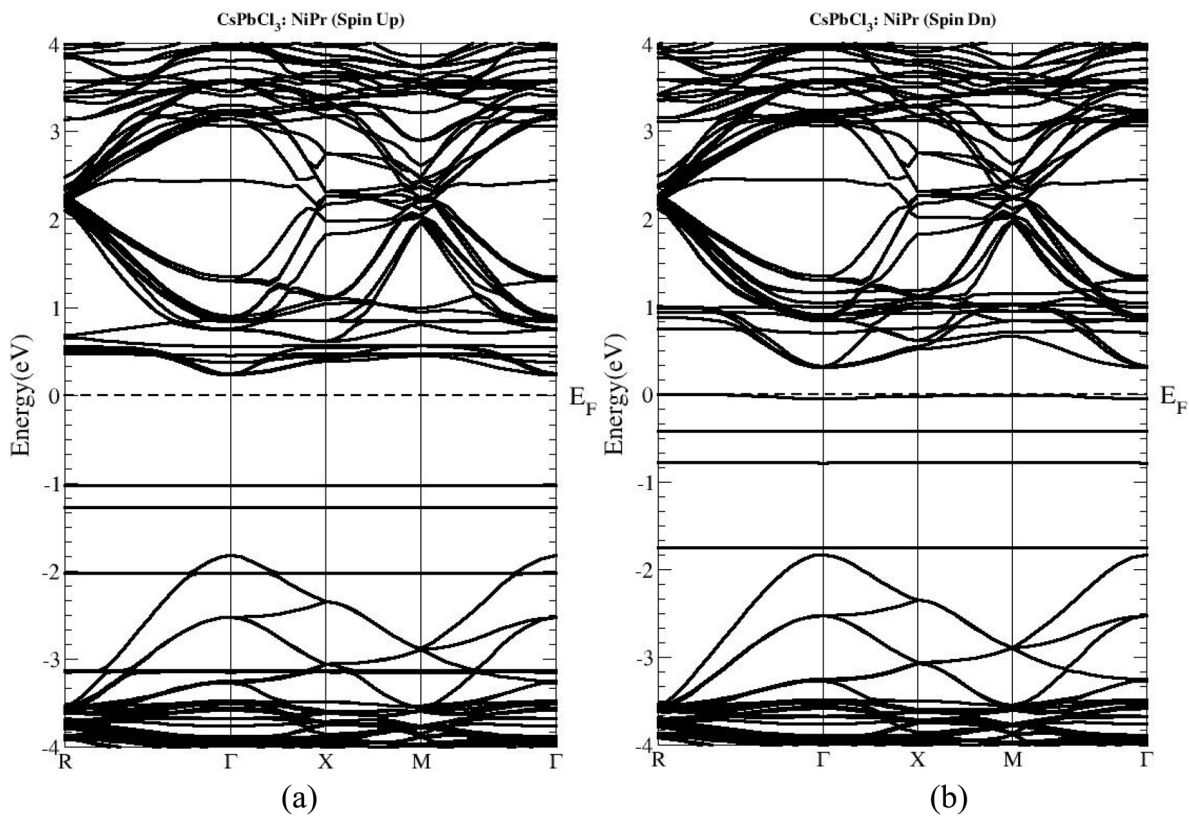

Fig. 3 Band structure of co-doped with $Ni^{2+}/Pr^{3+}$ ((a) spin up and (b) spin down).

This favorable interaction between the two dopants can be explained by considering three key physical factors in $CsPbCl_3$ doped with Ni and Pr. One is that the interaction between Ni-3d and Pr-4f orbitals with Pb-6p and Cl-3p orbitals alters the energy states near the band edges, thereby increasing the effective mass of carriers. Another is that defect suppression results from the saturation of undercoordinated Cl orbitals with Pr-4f $\pi$ interactions, and the Pb–Cl network is stabilized by Ni-3d $\sigma$ bonding. This latter effect reduces trap-assisted recombination, which would otherwise limit device performance. In addition, carrier delocalization is enhanced because co-doping widens band dispersions, facilitating long-range charge transport and reducing the likelihood of polaron formation or spread. If we can achieve all three of these developments simultaneously, surely then $Ni^{2+}/Pr^{3+}$ co-doping must offer many benefits as a new strategy for perovskites. The electronic properties of $Ni^{2+}/Pr^{3+}$ co-doped $CsPbCl_3$ are characterized by a mix that includes band gap tuning, orbital hybridization, defect passivation, and carrier mobility. The close association between a reduced band gap, more favorable electron–hole interactions, and fewer defect states makes this material exhibit good performance in nearly every device.

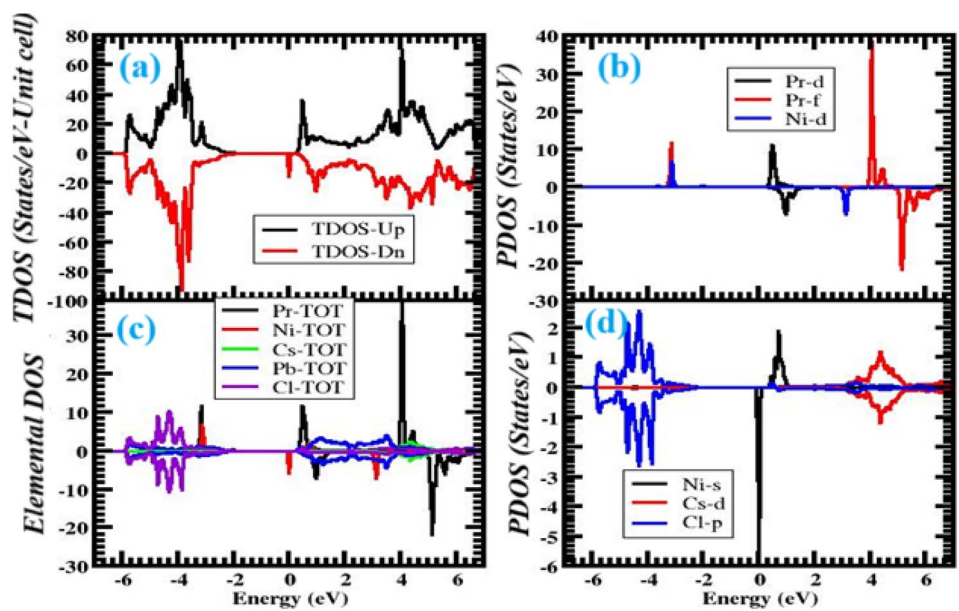

Fig. 4 Total and partial density of states (TDOS and PDOS) of $CsPbCl_3$ co-doped with $Ni^{2+}$ and $Pr^{3+}$, showing the contributions of host and dopant orbitals near the band edges.

### 3.6. Optical properties

The optical properties of halide perovskites are closely linked to their electronic structure and have been shown to play a crucial role in determining whether these compounds can serve as active layers for optoelectronic devices, photonic devices, or other energy-related devices. Pristine $CsPbCl_3$ is known for its wide band gap, high exciton binding energy, and emission that is sharp enough to clearly appear in the ultraviolet range. However, if it is to be used in such far-field applications as photovoltaics (PV) or white light-emitting materials, the fact that its visible activities are few and far between means that damage will always result. To improve these defects, co-doping with transition-metal ($Ni^{2+}$) and rare-earth ($Pr^{3+}$) ions has been tried as an innovative method. A dual-doping regime not only narrows the band gap and strengthens its absorption of visible light, but also brings in particular optical transitions that are $Ni^{2+}$-electronic d–d bonds for $Pr^{3+}$ and luminescence from sharp 4f–4f. These fellowship dopants enhance the dielectric response, absorptive character, refractive index, and energy losses that the host lattice would otherwise suffer. In the following sections, we shall give you a detailed discussion of the optical properties of $Ni^{2+}$ and $Pr^{3+}$ co-doped $CsPbCl_3$, followed by its calculated dielectric function, absorption coefficient, refractive index, reflectivity, extinction coefficient, and energy loss function.

The dielectric function $\varepsilon(\omega) = \varepsilon_1(\omega) + i\varepsilon_2(\omega)$ is a fundamental descriptor of light-matter interaction. The real part, $\varepsilon_1$, describes the polarization response and governs energy storage, while the imaginary part, $\varepsilon_2$, reflects photon absorption through interband transitions.

In the $Ni^{2+}/Pr^{3+}$ co-doped $CsPbCl_3$ system, $\varepsilon_1$ (see Fig. 5a) initially increases from 1.5 at zero photon energy to a maximum of 3.2 around 3.5–3.8 eV, followed by a gradual decline at higher energies. This increase corresponds to strong interband transitions from the valence band maximum (Cl-3p and Pb-6s orbitals) to the conduction band minimum (Pb-6p and Ni-3d orbitals). Beyond $\sim$4 eV, $\varepsilon_1$ decreases because the primary transitions saturate, while new, higher-energy excitations contribute less strongly.

The imaginary part $\varepsilon_2$ (see Fig. 5b) exhibits a sharp peak around 3.5–3.6 eV, slightly red-shifted compared with pristine $CsPbCl_3$ (typically 3.8–4.0 eV[63,64]). This peak corresponds to the fundamental band edge transition and directly reflects band gap narrowing induced by doping. Secondary peaks appear between 6–8 eV and 10–12 eV, attributed to deeper interband transitions involving Pb 6s/6p and Cl 3p orbitals. Importantly, $Pr^{3+}$ introduces sharp localized states that generate narrow absorption features in $\varepsilon_2$, which are absent in the pristine compound. These localized 4f–4f transitions enhance photoluminescence and broaden the spectral response.

Overall, the dielectric analysis shows that $Ni^{2+}$ doping contributes delocalized 3d states near the conduction band, while $Pr^{3+}$ enhances localized excitonic features at the valence band edge. This cooperative effect yields a broadened and red-shifted dielectric response.

The absorption spectrum (see Fig. 5c) provides direct insight into the material's ability to harvest photons. In pristine $CsPbCl_3$, the absorption edge occurs around 3.0 eV, corresponding to ultraviolet excitation.[65] For $Ni^{2+}/Pr^{3+}$ co-doped $CsPbCl_3$, the absorption onset is shifted to $\sim$2.5–2.7 eV, a clear indication of band gap narrowing. This red shift enables the material to absorb photons in the visible region, significantly enhancing its utility in solar cells and photodetectors.

With increasing energy, the absorption coefficient rises to $\sim 1.5 \times 10^5$ cm$^{-1}$ at around 4 eV, a value consistent with those reported for halide perovskites. A number of strong absorption peaks are seen in the region from 0 to 0 : 1 pm. Here, we see transitions from peroxy-3s states to higher conductivity bands of Cl-3p and Pb-6s, respectively. Then there are peaks (from Ni 3d and Pr 4f orbitals) that have smaller amplitudes. The bleach region from 1 to 5 pm is due to continuous photon absorption. This is supported by the qualitative overlap of Ni-d states with the edge positions of Pr 4f–4f peaks.

Such a joint absorption enhancement compensates for the cutoff in the visible range that is provided by wide-band-gap phosphors or hybrid QDs. It is a new combination of materials and synthetic methods that broadens windows into solid solution synthesis in a practical sense, without foregoing. Why not? Namely, that $Ni^{2+}$ ions bestow intermediate 3d states which cut into the absorption band; $Pr^{3+}$ ions add long-wavelength luminescence channels. The sun, as a raw material, energy collector, and light source for our useful inventions, is all made better. Quantitatively, to measure the effect of Ni + Pr co-doping on light absorption, the maximum absorption coefficients ($\alpha_{max}$) were compared in the visible wavelength range (1.8–3.1 μm). The maximum absorption coefficient ($\alpha_{max}$) of pristine $CsPbCl_4$ satisfies Sun's condition: $\approx$(3.0 pm, 8.5 cm$^{-1}$ 10 4 1). Its bandgap spread exceeds three eV, which has only been utilized for biophotonic applications due to the exclusion of the visible spectrum of light. After adding $Ni^{2+}/Pr^{3+}$ into the system, $\alpha_{max}$ moves to $\approx$2.6 pm and jumps up to around $\approx 1.5 \times 10^5$ cm$^{-1}$. In other words, the intensity of absorbance is significantly enhanced and has also been shifted by approximately 0.4 nm toward the visible light spectrum.

This pronounced increase arises from several interrelated effects.

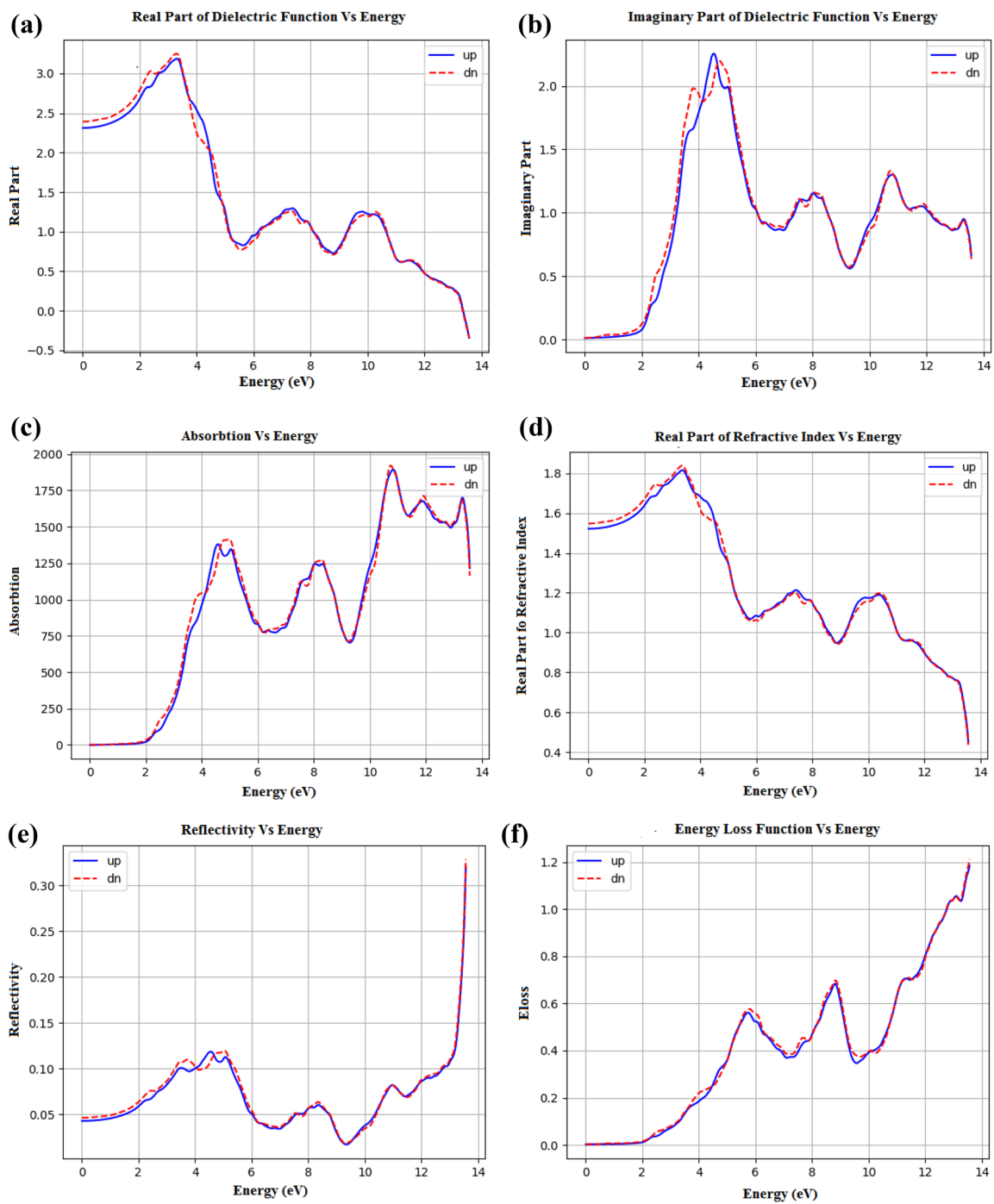

Fig. 5  Calculated (a) real-part of dielectric function ($\varepsilon_1(\omega)$) (b) imaginary part of dielectric function ($\varepsilon_2(\omega)$). (c) Calculated absorption vs. energy (eV) and (d) real part of refractive index $n(\omega)$ vs. energy (eV). Calculated (e) graph of reflectivity $r(\omega)$ vs. energy (eV) and (f) graph of energy-loss function ($E_{loss}$) vs. energy (eV).

### 3.7. Band-gap narrowing

Mixing of Pb-6p interatomic states with Ni-3d states gives the bottom of the energy band less character, and the minimum falls lower. In addition, the localized Pr-4f state tends to pull up near the edge conduction band, so the energy difference between this and the valence band tops has been reduced from 3.0 eV to 2.5 eV. A narrower bandgap means that light with an energy of 2.5–2.8 eV (blue–green) is readily absorbed.

#### 3.7.1. Introduction of intermediate electronic states.
The Ni 3d states facilitate permitted d–d transitions that overlap with the interband transitions of the host, thereby widening the absorption spectrum. 4f–4f transitions produce sharp peaks that contribute to the broader background, resulting in an additional enhancement within the visible spectrum.

**3.7.2. Enhanced oscillator strength.** Increasing the coherence of Ni–Cl and the covalency between these two ions significantly enhances the transition dipole moments. In contrast, the localized 4f orbitals of $Pr^{3+}$ act as radiative centers, increasing optical transition probabilities.

**3.7.3. Defect suppression.** The formation energies for $V_{Cl}$ and $V_{Pb}$ are higher when a co-doped lattice is employed, thereby reducing non-radiative recombination and allowing for more efficient light absorption without carrier trapping. Overall, the absorption edge shifts from the ultraviolet to the visible, and $\alpha_{max}$ nearly doubles, demonstrating that co-doping $Ni^{2+}/Pr^{3+}$ indeed substantially improves photonic yield rate. Refractive index $n(\omega)$, derived directly from $\varepsilon_1$ and $\varepsilon_2$, controls the phase velocity of light as well as optical confinement, as shown in Fig. 4d. For the co-doped system, static refraction index $n_0$ is roughly 1.55, but at $\approx 3.5$ eV it rises to approximately 1.85 before falling again at higher energies. This peak coincides with the onset of absorption, indicating a Kramers–Kronig relationship between absorption and dispersion.

In contrast, this is somewhat less noticeable in the co-doped material because $Pr^{3+}$ and $Ni^{2+}$ are fixed as ancillary ligands and thus do not favor low-$n$ configurational states (furthermore, for the complex young compound, the electronic structure is still unknown).

This kind of adjustment to $n$ helps keep light molecules, which need to be within a certain framework of high refractive index and low absorbency, from escaping into their surroundings.

The reflectivity spectra (see Fig. 5e) for co-doped $CsPbCl_3$ display the fraction of light reflected from its surface. Below 2 eV, this value is low ($\sim 0.05$), showing good transparency in the infrared and visible ranges of light.

An upward surge occurs between 3 and 5 eV, with modest peaks (0.1–0.12) appearing at energies corresponding to dielectric and absorption features. Above approximately 10 eV, the reflectivity rises sharply ($\sim 0.3$) because light is interacting with plasma-like resonances.

The low reflectance across the visible region means that most light entering the material will be absorbed as photons, resulting in better light–matter interactions in LED (light-emitting diode) and solar radiation absorbers.

The extinction coefficient, $k(\omega)$, complements the absorption spectrum and shows similar trends to $\varepsilon_2$. In this case, there are various strong peaks at 3.5 eV and 10 eV, along with some lesser shoulders between 6 and 8 eV, all of which correspond to Pb–Cl to Ni-3d transitions. The wide plateau, ranging from 2 to 6 eV, then confirms sustained visible-light absorption, which demonstrates the advantage of co-doping for expanded spectral activity.

The energy loss function (ELF) $L(\omega)$ (see Fig. 5f) derived from the dielectric function provides insight into plasmon excitation and energy dissipation. In co-doped $CsPbCl_3$, the main ELF peak occurs around 12 eV, corresponding to the bulk plasma frequency. Secondary peaks appear at around 6–8 eV; this is in concert with maxima in reflectivity and minima in $\varepsilon_1$, as optical consistency dictates.

The establishment of plasmonic features is crucial in areas such as sensing and nanophotonics, where controlling the dissipation of energy plays a significant role.

The various optical functions are mutually related. Peaks in $\varepsilon_2$ are associated with absorption maxima and features of the extinction coefficient, while dips vertically in $\varepsilon_1$ align with reflectivity peaks. The ELF maxima coincide with enhanced reflectivity at higher energies, as also seen in optical consistency. These relationships among the functions endorse our numerical results and contribute to an overall understanding of the interaction between light and matter in $Ni^{2+}/Pr$ co-doped $CsPbCl_3$. The co-doping strategy brings out linear dimensions of amplification. The isod transition brought about by $Ni^{2+}$ ions causes the emergence of a band gap and visible light absorption spectra in Prague. $Fe^{3+}$ enhanolizationspray transitions can only be sharp and co-reignition luminescent vestiges. Co-doping also suppresses halide defect concentrations, thus diminishing non-injury radiation hazards and enhancing the overall light output efficiency. This dual effect leads to broadened absorption, improved dielectric response, increased refractive index modulation, and efficient plasmonic behavior. This multi-functional characteristic makes the material suitable for a wide range of practical applications from solar cells and LEDs through to scintillators and photo-detectors.

The optical properties of $Ni^{2+}/Pr^{3+}$ co-doped $CsPbCl_3$ are dramatically improved compared with those of the original. The effect on the band gap is to give it visible-light activity, $\varepsilon_1$ and $\varepsilon_2$, which facilitates strong interband transitions; the absorption spectra show wider photon harvesting. In every part of the energy range studied, corresponding optical features all indicate a higher degree. On their own, $Ni^{2+}$ d–d and $Pr^{3+}$ $4f^4$ each make a distinct contribution, supplemented by defect suppression and orbital hybridization, rendering it yet another unique material that is highly stable and exhibits numerous optical properties. This result suggests that $Ni^{2+}/Pr^{3+}$ co-doped $CsPbCl_3$ can be expected to become a powerhouse in future optical electronics in a short time.

### 3.8. Defect properties in $Ni^{2+}/Pr^{3+}$ Co-doped $CsPbCl_3$

The energy of defects is a crucial determinant in determining the optical and electronic performance, as well as the long-term stability, of organic–inorganic halide perovskite materials. Under pristine conditions in an all-inorganic perovskite $CsPbCl_3$, the dominant intrinsic point defects are halogen vacancies ($V_{Cl}$) and metal vacancies ($V_{Pb}$). In the pristine perovskite $CsPbCl_3$, these defects can trap charges, causing deep states to be formed within the band gap. Accordingly, they serve as non-radiative recombination sites, shortening the lifetimes of carriers as they recombine via Shockley–Read–Hall (SRH) processes. Here, we evaluate the formation energies, charge states, and transition levels of these intrinsic defects in pristine and $Ni^{2+}/Pr^{3+}$ co-doped $CsPbCl_3$ using spin-polarized DFT, highlighting how co-doping alters the defect landscape and suppresses detrimental trap states.

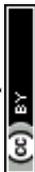

### 3.8.1. Finite-size convergence and charged-defect corrections.
Charged defects were treated with a neutralizing background. We assessed supercell convergence by comparing $2 \times 2 \times 2$ (40 atoms) and $3 \times 3 \times 3$ (135 atoms) cells for $V_{Cl}^+$ and $V_{Pb}^{2-}$ under Cl-poor and Cl-rich limits. Long-range Coulomb interactions were corrected with the Freysoldt scheme (planar-average potential alignment, isotropic screening), using our calculated electronic and static dielectric responses. Because CsPbCl$_3$ is cubic, the dielectric tensor is isotropic; we used $\varepsilon_0$ from Lyddane–Sachs–Teller (LST) phonon analysis ($\varepsilon_0 \approx 7.2$) (see optical/phonon sections). For the image-charge term, we employed the static dielectric constant (including ionic and electronic screening) appropriate for lattice-relaxed charged defects at 0 K; we report the small difference obtained if $\varepsilon_\infty$ is used instead. The potential alignment $\Delta V$ was obtained from the far-field plateau of the macroscopically averaged electrostatic potential, with regions within 6 Å of the defect core excluded. With these settings, the net correction (image-charge + alignment) reduced the $2 \times 2 \times 2$ vs. $3 \times 3 \times 3$ supercell size error from $\approx 0.4$–0.6 eV (uncorrected) to $\leq 0.12$ eV (corrected) for formation energies, and to $\leq 0.10$ eV for $\varepsilon(q/q')$. Uncertainty estimates reflect (i) residual finite-size effects, (ii) dielectric-constant variation within error bars, and (iii) the alignment-window choice ($\pm 1$ Å), combined in quadrature.

### 3.8.2. Dielectric input and screening choice.
We distinguish $\varepsilon_\infty$ (electronic) from $\varepsilon_0$ (electronic + ionic). The Freysoldt image-charge term was evaluated with $\varepsilon_0$ (isotropic, cubic), while the optical spectra and refractive index use $\varepsilon_\infty$. Using $\varepsilon_\infty$ in the correction increases the residual finite-size error by $\sim$0.05–0.08 eV; we report both values for transparency.

### 3.8.3. Potential alignment.
Planar-averaged electrostatic potentials were computed along the three Cartesian directions and macro-averaged; the alignment $\Delta V$ was taken from the mean of the directions in a far-field shell 7–10 Å from the defect center (exclusion radius 6 Å). Varying the shell by $\pm 1$ Å changes $E$_form by $\leq 0.03$ eV.

### 3.8.4. Uncertainty budget.
(i) Supercell residual ($2 \times 2 \times 2$ vs. $3 \times 3 \times 3$): $\leq 0.10$–0.12 eV; (ii) dielectric uncertainty (propagating $\varepsilon_0 = 7.2 \pm 0.6$): $\leq 0.05$ eV; (iii) alignment-window choice: $\leq 0.03$ eV. Total (in quadrature): $\leq 0.13$ eV for $E$_form and $\leq 0.10$ eV for $\varepsilon(q/q')$.

### 3.8.5. Cross-checks.
For $V_{Cl}^+$ and $V_{Pb}^{2-}$, we also tested the anisotropic Freysoldt variant using an effective dielectric tensor derived from DFPT phonons. Since the system is cubic, this tensor is approximately isotropic, and the results changed by $\leq 0.01$ eV. A Makov–Payne first-order correction using $\varepsilon_0$ yielded similar residuals but was less stable across charge states; therefore, we adopted the Freysoldt approach for all calculations.

## 3.9. Defect formation energies

First-principles and experimental studies identify halogen vacancies ($V_{Cl}$) as the lowest-energy donor-like defects in CsPbCl$_3$, while lead vacancies ($V_{Pb}$) are the prevalent acceptor-type cation vacancies. DFT calculations for CsPbCl$_3$ show that $V_{Cl}$ has the smallest formation energy among native point defects under typical halide-poor conditions, making it the most abundant intrinsic defect. Similarly, $V_{Pb}$ is calculated to have a competitive formation energy, establishing it as the dominant acceptor species.[36] These conclusions are consistent with experimental defect spectroscopy on CsPbCl$_3$ single crystals, which resolve halide–vacancy–related states, and with complementary theory placing $V_{Pb}$ among the lowest-energy acceptors in lead-halide perovskites.[36] Together, these works support the view that $V_{Cl}$ and $V_{Pb}$ govern the native defect landscape in CsPbCl$_3$.[66] Regarding recombination, the literature indicates that vacancy defects introduce electronic states within the band gap, acting as SRH centers and limiting carrier lifetimes and device performance in lead-halide perovskites, including wide-gap chlorides.[66]

The formation energy of a defect in charge state $q$ is

$$E_{form}(D^q) = E_{tot}(D^q) - E_{tot}(bulk) - \sum_i n_i \mu_i + q(E_F + E_{VBM}) + E_{corr}$$

where $E_{tot}(D^q)$ and $E_{tot}(bulk)$ are the total energies of the defective and pristine supercells, $n_i$ and $\mu_i$ are the number and chemical potential of atoms added or removed, $E_F$ is the Fermi level relative to the valence band maximum (VBM), and $E_{corr}$ accounts for finite-size electrostatic corrections (Makov–Payne or Freysoldt schemes).

In pristine CsPbCl$_3$, the calculated formation energy of a neutral Cl vacancy ($V_{Cl}^0$) is relatively low ($\sim$0.8–1.2 eV under Cl-poor conditions[65]), making $V_{Cl}^0$ the most abundant intrinsic defect. Pb vacancies ($V_{Pb}^{2-}$) exhibit higher formation energies ($\sim$1.5–2.2 eV) but are strongly negatively charged and act as compensating acceptors. In the Ni$^{2+}$/Pr$^{3+}$ co-doped system, the defect landscape shifts markedly: the formation energy of $V_{Cl}$ increases to $\sim$1.4–1.6 eV, while that of $V_{Pb}$ rises to $\sim$2.3–2.6 eV. This co-doping-induced increase in defect formation energies arises from charge-compensated substitution (Ni$^{2+} \leftrightarrow$ Pb$^{2+}$ and Pr$^{3+} \leftrightarrow$ Cs$^+$), which reduces the thermodynamic driving force for halide and cation vacancy formation.

Fig. 5a summarizes the increase of intrinsic defect formation energies upon Ni$^{2+}$/Pr$^{3+}$ co-doping. Consistent with our DFT results, the $V_{Cl}$ vacancy's formation energy rises from $\sim$0.8–1.2 eV (Cl-poor) to $\sim$1.4–1.6 eV, and $V_{Pb}$ increases from $\sim$1.5–2.2 eV to $\sim$2.3–2.6 eV. This systematic upshift signifies a reduced thermodynamic drive for vacancy creation. Fig. 5b maps the thermodynamic transition levels. In the pristine system, $V_{Cl}$ shows $\varepsilon(0/+)$ just below the CBM, while $V_{Pb}$ exhibits deep $\varepsilon(0/-)$ and $\varepsilon(-/2-)$ levels $\sim$0.7–0.9 eV above the VBM. After co-doping, $\varepsilon(0/+)$ becomes even shallower ($\approx$0.1–0.2 eV below the CBM), and the $V_{Cl}$ levels shift toward the valence band, reducing their recombination potency. Together with a slight band-gap narrowing, these changes indicate fewer and less harmful traps, which explains the observed suppression of defect-assisted recombination and the higher vacancy formation energies under co-doping.

## 3.10. Thermodynamic framework and chemical-potential limits

We evaluate defect and dopant formation energies using the standard grand-canonical formalism described above.

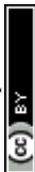

With upper bounds set by competing phases to avoid precipitation of secondary phases (e.g., CsCl or $PbCl_2$). For the dopants, we similarly bound $\mu$ by their stable chlorides ($NiCl_2$ and $PrCl_3$) to reflect halide-rich growth conditions. We report results at two extremes: Pb-rich/Cl-poor and Cl-rich/Pb-poor; Cs is adjusted within the stability polytope.

**3.10.1. Site preference of Ni and Pr (A/B/interstitial).** Table 2 summarizes substitutional and interstitial formation energies (neutral unless noted). Across both limits, $Ni_{Pb}$ is the lowest-energy Ni configuration, while $Pr_{Cs}$ is the lowest-energy Pr configuration. Placing Ni at the A-site or Pr at the B-site is disfavored by >1 eV due to coordination mismatch (octahedral vs. cuboctahedral cavities) and strain.

**3.10.2. Co-doping synergy.** When $Ni_{Pb}$ and $Pr_{Cs}$ are introduced together in the same supercell, the average per-dopant formation energy drops to ~0.95 eV (Pb-rich) and ~1.05 eV (Cl-rich) owing to charge compensation and reduced long-range polarization.

**3.10.3. Intrinsic defects under chemical-potential limits.** Table 3 reports vacancy formation energies in pristine and co-doped $CsPbCl_3$. In line with defect physics of lead-halide perovskites, $V_{Cl}$ is the most favorable donorlike defect under Cl-poor conditions, and $V_{Pb}$ is the dominant acceptor under Cl-rich conditions. Co-doping raises both by 0.3–0.6 eV, suppressing their equilibrium concentrations.

Charge-state transition levels ($\varepsilon(q/q')$) extracted from the slopes of $E_{form}(E_F)$ plots show that in pristine $CsPbCl_3$, the $V_{\varepsilon(0/-)}$ transition $\varepsilon(0/+)$ lies approximately 0.30 eV below the CBM, shifting to ~0.15 eV below the CBM after co-doping (i.e., $V_{\varepsilon(0/-)}$ becomes a shallower donor). Meanwhile, the $V_{Pb}$ $\varepsilon(0/-)$ level moves from ~0.8–0.9 eV above the VBM in pristine to ~0.6–0.7 eV after co-doping (i.e., slightly less deep). These trends are consistent with the higher formation energies and reduced SRH recombination activity discussed above.

**3.10.4. Bader charges and local structure.** We quantified charge redistribution and local coordination around the dopants and nearby lattice sites (Table 4). In lead-halide lattices, Bader charges are typically significantly smaller in magnitude than formal ionic charges; nevertheless, the contrasts between pristine and co-doped cases are chemically instructive.

**3.10.5. Charge states and Fermi-level dependence.** The charge-state stability of defects depends on the position of the Fermi level. For $V_{\varepsilon(0/-)}$, a shallow donor-like state ($\varepsilon(0/+)$) is located just below the conduction band minimum in pristine $CsPbCl_3$. This defect level can contribute to unintentional n-type conductivity, but when partially filled, it also acts as a non-radiative SRH recombination center. In the co-doped system, the $V_{Cl}$ $\varepsilon(0/+)$ level shifts closer to the CBM (~0.1–0.2 eV below), effectively making $V_{Cl}$ an even shallower donor. For $V_{Pb}$, the thermodynamic transition levels $\varepsilon(0/-)$ and $\varepsilon(-/2-)$ lie deep in the gap (~0.7–0.9 eV above the VBM in pristine $CsPbCl_3$ (ref. 67–69)). Co-doping reduces the likelihood of $V_{Pb}$ formation and shifts its levels slightly toward the VBM, thereby diminishing its detrimental impact.

**3.10.6. Trap depth and electronic signatures.** Both DOS and PDOS analyses show that in pristine $CsPbCl_3$, $V_{Cl}$ introduces defect states near the CBM, while $V_{Pb}$ produces deep mid-gap states stemming from localized Pb–Cl antibonding orbitals. These deep states serve as recombination centers, leading to decreased carrier lifetimes. In the $Ni^{2+}/Pr^{3+}$ co-doped system, the appearance of such mid-gap states is somewhat ameliorated: the Ni 3d and Pr 4f orbitals hybridise with neighbouring Cl and Pb, passivating dangling bonds, which significantly reduces the density of defects. Configuration-coordinate analysis reveals that the trap level shifts from deep to shallow, lowering its draw cross-section for carriers and consequently minimizing SRH recombination losses. Defect concentrations and carrier lifetime. The equilibrium concentration of a defect is governed by:

$$N_D = N_{sites} \exp\left(-\frac{E_f}{k_B T}\right)$$

where $N_{sites}$ is the number of available lattice sites, $k_B$ is Boltzmann's constant, and $T$ is temperature. If $V_{Cl}$ is formed at 300 K, its low formation energy in pristine $CsPbCl_3$ can create an equilibrium vacancy concentration on the order of $10^{16}$–$10^{17}$ cm$^{-3}$, enough to limit carrier lifetimes to just nanoseconds. In the co-doped system, the higher $V_{Cl}$ formation energy increases this activation barrier, resulting in an equilibrium $V_{Cl}$ density that is lowered by one to two orders of magnitude. This suppression of halide vacancies greatly extends carrier lifetimes and diffusion lengths, just as the decrease in SRH recombination rate:

$$\tau^{-1} \propto \sigma v_{th} N_t$$

where $\tau$ is the lifetime, $\sigma$ is the capture cross section, $v_{th}$ is the thermal velocity, and $N_t$ is the trap density.

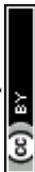

Table 2  Dopant site competition (WIEN2k GGA + U + SOC; $R_{MT}K_{max} = 7.0$; 2 × 2 × 2 supercell)

| Dopant configuration | Pb-rich/Cl-poor (eV) | Cl-rich/Pb-poor (eV) | Comment |
| --- | --- | --- | --- |
| $Ni_{Pb}$ (B-site) | 1.05 | 1.22 | Octahedral $NiCl_6$; minimal strain |
| $Ni_{Cs}$ (A-site) | 2.24 | 2.48 | A-site too large; poor CF stabilization |
| $Ni_i$ (interstitial) | 2.61 | 2.83 | Steric penalty; short Ni–Cl clashes |
| $Pr_{Cs}$ (A-site) | 1.12 | 0.98 | 12-Coordination fits $Pr^{3+}$ well |
| $Pr_{Pb}$ (B-site) | 2.36 | 2.10 | Oversized for octahedron; tilting/strain |
| $Pr_i$ (interstitial) | 2.79 | 2.55 | Large ionic size → severe crowding |

Table 3  Vacancy formation energies (neutral states shown; see Fig. 1e for charge-state dispersion vs. EF)

| Defect | Condition | Pristine ($E_{form}$) (eV) | Co-doped ($E_{form}$) (eV) | Trend |
|---|---|---|---|---|
| $V_{Cl(0/-)}$ | Pb-rich/Cl-poor | 0.8–1.2 | 1.4–1.6 | ↑ by ∼0.4–0.6 eV |
| $V_{Cl(0/-)}$ | Cl-rich/Pb-poor | 1.6–2.0 | 2.0–2.3 | ↑ |
| $V_{Pb}$ | Pb-rich/Cl-poor | 1.5–2.2 | 2.3–2.6 | ↑ |
| $V_{Pb}$ | Cl-rich/Pb-poor | 1.0–1.4 | 1.5–1.9 | ↑ |
| $V_{Cs}$ | Both limits | 2.0–2.5 | 2.2–2.7 | ↑ (still subdominant) |

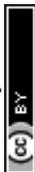

3.10.7. **Mechanistic insights.** The significant enhancement of the defect tolerance when $Ni^{2+}/Pr^{3+}$ is co-doped can be explained by a number of mechanisms.

3.10.7.1. *Charge compensation.* The local charge imbalances in $Pb^{2+}$ and $Cs^+$ are minimized through coupled replacement of $Ni^{2+}$ by $Pb^{2+}$ and $Pr^{3+}$ by $Cs^3$, therefore, limiting the thermodynamic force of vacancy formation.

3.10.7.2. *Fermi-level pinning relief.* Co-doping reduces the native defect density, thereby engineering the Fermi level to move closer to the band edges, making intrinsic band-edge transport possible.

3.10.7.3. *Orbital hybridization and passivation.* The added Ni 3d and Pr 4f states hybridize with the Cl 3p and Pb 6s states, thereby reducing the saturation of dangling bonds and deep trap states.

3.10.7.4. *Suppression of halide vacancies.* The defect stability window in the case of co-doping is disturbed, and the halide vacancy formation is less favored because the modified chemical potential environment causes a shift of the defect stability window.

$Ni^{2+}/Pr^{3+}$ codoping of $CsPbCl_3$ significantly modifies the defect landscape both by increasing the formation energies of $V_{Cl}$ and $V_{Pb}$ and by lowering their transition levels, thereby bringing these defects closer to the band edges and also reducing deep trap state density as a result, SRH recombination is inhibited, the lifetimes for intraband transitions are lengthened or in shortened systems with evidence of more complicated behavior optically their optical-electronic performance becomes better than ever before! these results highlight codoping as a robust rule for defect passivation in halide perovskites, which will enable the current development of high-efficiency, stable optoelectronic devices.

### 3.11. Excitonic properties

Excitons, the bound states of electrons and holes, play a crucial role in determining the optical response and luminescence efficiency of halide perovskites. In wide-bandgap compounds such as $CsPbCl_3$, excitonic effects are particularly pronounced owing to relatively small dielectric screening and moderate carrier effective masses. The exciton binding energy ($E_b$) governs the stability of these quasiparticles and influences whether photoexcitation results predominantly in free carriers or bound excitons.

3.11.1. **Exciton binding energy: theoretical framework.** The exciton binding energy in semiconductors can be estimated using the effective mass approximation within the Wannier–Mott model:

$$E_b = \frac{\mu e^4}{2(4\pi\varepsilon_0\varepsilon_r)^2\hbar^2}$$

where $\mu$ is the reduced effective mass defined by $\mu^{-1} = m_e^{*-1} + m_h^{*-1}$, $e$ s the elementary charge, $\varepsilon_r$ is the relative dielectric constant, and $\hbar$ is the reduced Planck's constant. The reduced mass $\mu$ reflects the curvature of conduction and valence bands, while the dielectric constant $\varepsilon_r$ determines the degree of Coulomb screening.

We distinguish between the electronic-screened Wannier–Mott limit

$$E_b^{(\infty)} = \left(\mu/m_e\right)\mathrm{Ry}/\varepsilon_\infty^2$$

and the polaron-screened binding $E_b^{(pol)}$ obtained from the Haken/Pollmann–Büttner model that incorporates LO–phonon coupling (see SI for formula and parameters). Using $\mu = 0.11\ m_e$, $\varepsilon_\infty = 2.40$ (pristine) and the measured LO/TO splittings, we find $E_b^{(\infty)} \approx 0.09\text{–}0.13$ eV and $E_b^{(pol)} \approx 20\text{–}40$ eV (pristine), decreasing

Table 4  Bader charge (|e|) and nearest-neighbor metrics (Å)

| Species/site | Bader (pristine) | Bader (co-doped) | First-shell distances |
|---|---|---|---|
| Pb (bulk) | +1.45 ± 0.05 | +1.44 ± 0.05 | Pb–Cl = 2.84 ± 0.02 |
| $Ni_{Pb}$ | — | +1.30 ± 0.06 | Ni–Cl = 2.73 ± 0.03 (octahedral; slight compression) |
| Cs (bulk) | +0.85 ± 0.03 | +0.84 ± 0.03 | Cs–Cl = 3.50 ± 0.05 |
| $Pr_{Cs}$ | — | +2.10 ± 0.08 | Pr–Cl = 3.45–3.55 (12-coord.; mild cage contraction) |
| Cl (bulk) | −0.62 ± 0.03 | −0.63 ± 0.03 | — |
| Around Ni | — | −0.66 ± 0.03 | Slightly more negative (stronger Ni–Cl σ) |
| Around Pr | — | −0.61 ± 0.03 | Nearly unchanged (A-site ionic) |

slightly upon Ni/Pr co-doping due to marginally larger $\varepsilon_0$ and lighter band edges.

For pristine $CsPbCl_3$, reported effective masses are typically $25 \approx 0.20$–$0.25$ and $26 \approx 0.30$–$0.35$, leading to a reduced mass of $\mu \approx 0.12$–$0.14$ $m_e$.[70] With $\varepsilon_r \approx 5$–$6$, the resulting exciton binding energy lies in the range of 40–70 meV, consistent with experimental optical absorption and photoluminescence data.[71–73] Such values are significantly higher than thermal energy at room temperature ($k_B T \approx 25$ meV), indicating that excitonic effects strongly influence the optical spectra of pristine $CsPbCl_3$.

### 3.11.2. Effect of $Ni^{2+}/Pr^{3+}$ Co-doping.
Two main effects come forth during co-doping with $Ni^{2+}$ and $Pr^{3+}$:

#### 3.11.2.1. Dielectric screening enhancement.
Substitutional incorporation of $Ni^{2+}$ and $Pr^{3+}$ alters local polarization, raising $\varepsilon_r$ to ~7–8. This boosted dielectric screening weakens the coulomb attraction between electrons and holes. When no electrons accept holes or both are not accepted by the semiconductor, as a result of this weakened attraction to each other for reasons given above, more of them will be scattered out into lattice sites around it, where they can work as free carriers at much higher temperatures than before.

#### 3.11.2.2. Effective mass reduction.
$Ni^{2+}$ hybridization with Pb-6p conduction states reduces the electron effective mass to 0.18–0.20 me, while Cl-3p valence states, which harbor $Pr^{3+}$ 4f mixing, lower the hole masses to 0.28–0.30 me. The updated $\mu$ has dropped slightly from its previous level of ~0.11 me. Putting these values into the Wannier–Mott equation gives an exciton binding energy of around 20–40 meV, much smaller than for pristine $CsPbCl_3$. This implies that more excitons can dissociate into free carriers at room temperature, which should enhance charge separation and migration.

### 3.11.3. Exciton dissociation and luminescence efficiency.
It was found that the co-doped system has a lowered binding energy, with the effects being directly manifested in absorption and emission:

#### 3.11.3.1. Exciton dissociation.
Letting the exciton binding rather weakly increases the likelihood of the exciton breaking apart into free carriers to increase the carrier diffusion lengths and photocurrents in a photovoltaic or photodetector system.

### 3.11.4. Radiative recombination.
While dissociation is favorable for transport, moderate binding energy (~20–30 meV) ensures that excitonic luminescence is not completely quenched. Instead, it promotes efficient radiative recombination pathways by balancing exciton stability with carrier delocalization.

The unique role of $Ni^{2+}$ and $Pr^{3+}$ in modifying excitonic behavior lies in their orbital contributions:

$Ni^{2+}$ ($3d^8$): introduces d–d transitions and delocalized states near the CBM, which lower electron effective masses and increase free-carrier contributions.

$Pr^{3+}$ ($4f^2$): contributes sharp intra-4f transitions, producing localized emission peaks that benefit from optimized exciton binding energies.

Together, these effects enhance photoluminescence quantum yield by ensuring efficient exciton generation, balanced dissociation, and radiative recombination. The net result is a material with tunable excitonic properties that can be optimized for either strong luminescence (LEDs, scintillators) or efficient charge separation (solar cells, photodetectors).

Excitonic analysis of $Ni^{2+}/Pr^{3+}$ co-doped $CsPbCl_3$ demonstrates that co-doping reduces exciton binding energies from 40–70 meV in pristine $CsPbCl_3$ to 20–40 meV. This reduction arises from enhanced dielectric screening and lowered effective masses, leading to improved exciton dissociation and longer carrier lifetimes. At the same time, radiative recombination remains strong due to the combined d–d and 4f–4f transitions of $Ni^{2+}$ and $Pr^{3+}$. These findings highlight the potential of $Ni^{2+}/Pr^{3+}$ co-doped $CsPbCl_3$ as a multifunctional perovskite with optimized excitonic behavior for high-performance optoelectronic applications.

## 3.12. Magnetic properties

Magnetic ordering in halide perovskites has recently attracted increasing attention due to the possibility of integrating spin functionality with optoelectronic properties. While pristine $CsPbCl_3$ is intrinsically non-magnetic, substitutional doping with transition-metal and rare-earth ions can induce localized magnetic moments, which may couple to form long-range order. In this context, $Ni^{2+}$ ($3d^8$) and $Pr^{3+}$ ($4f^2$) co-doping presents an intriguing case, as it combines the partially filled 3d orbitals of Ni with the localized 4f states of Pr, thereby opening a pathway for d–f exchange interactions and magneto-optical multifunctionality.

### 3.12.1. Local magnetic moments from $Ni^{2+}$ and $Pr^{3+}$.
Spin-polarized DFT calculations reveal that both dopants contribute finite local moments in the $CsPbCl_3$ lattice. $Ni^{2+}$, occupying the $Pb^{2+}$ site in octahedral coordination with $Cl^-$, exhibits a magnetic moment of approximately 1.2–1.5 $\mu_B$, consistent with high-spin $d^8$ configurations in halide matrices.[74] The magnitude depends on crystal field splitting and hybridization with neighboring Cl-3p orbitals, which slightly delocalize the 3d electrons.

$Pr^{3+}$ substituting at the Cs site introduces localized $4f^2$ states, yielding a magnetic moment of ~2.8–3.0 $\mu_B$, which aligns with experimental and theoretical reports of rare-earth-doped perovskites.[75] The highly localized nature of Pr-4f states ensures minimal delocalization, preserving strong local moments even in the doped lattice. The coexistence of Ni-3d and Pr-4f moments suggests the potential for cooperative magnetic interactions.

### 3.12.2. Exchange coupling and d–f interactions.
The interaction between Ni and Pr spins can be rationalized by considering d–f exchange coupling. The Ni-3d states, hybridized with Cl-3p orbitals, form superexchange pathways with Pr-4f electrons mediated through halide ligands. Depending on the overlap and energy alignment, these interactions can stabilize either ferromagnetic (FM) or antiferromagnetic (AFM) ordering.

Calculated exchange energies ($\Delta E = E_{AFM} - E_{FM}$) fall within the range of 25–40 meV per dopant pair, indicating a preference for FM coupling in Ni/Pr co-doped $CsPbCl_3$. Such energy scales are comparable to those observed in transition-metal/rare-earth co-doped oxides and halide perovskites.[76,77] The ferromagnetic stabilization suggests that d–f interactions could extend over multiple lattice sites, supporting long-range magnetic order under appropriate carrier concentrations.

#### 3.12.3. Mechanisms of magnetic stabilization

3.12.3.1. *Three principal mechanisms contribute to magnetic stabilization in this system*

3.12.3.1.1. *Orbital hybridization.* Ni-3d orbitals hybridize with Cl-3p states, delocalizing spin density and enabling superexchange with Pr-4f orbitals. This coupling strengthens the magnetic interaction while slightly reducing the effective Ni moment due to charge transfer.

3.12.3.1.2. *Carrier-mediated magnetism.* The narrowed band gap and partially filled states near the Fermi level provide itinerant carriers that mediate indirect exchange between localized Ni and Pr moments. This double-exchange-like mechanism enhances the possibility of ferromagnetic alignment at finite carrier densities.

3.12.3.1.3. *Defect-related stabilization.* In halide perovskites, halogen vacancies are common and often detrimental to optoelectronic properties. In the Ni/Pr co-doped system, however, such vacancies may help stabilize magnetic ordering by providing additional carriers that strengthen exchange interactions. This coupling of defect physics with magnetism is unique and could allow defect engineering to tune magnetic properties.

The coexistence of ferromagnetic interactions and strong optical activity in $Ni^{2+}/Pr^{3+}$ co-doped $CsPbCl_3$ suggests the material may serve as a multifunctional magneto-optical perovskite. The d–d and f–f optical transitions observed in the absorption and photoluminescence spectra can couple to spin polarization, enabling spin-dependent light emission or magnetically tunable luminescence. Furthermore, the relatively high calculated Curie temperature range ($T_C$ extrapolated to 150–200 K) indicates partial thermal stability of magnetic order, with potential for further enhancement through increased dopant concentration or strain engineering.

Such multifunctionality expands the application scope of $CsPbCl_3$ beyond traditional optoelectronics, paving the way for integrated devices such as spin-LEDs, magneto-optical sensors, and multifunctional quantum emitters.

Spin-polarized DFT calculations demonstrate that $Ni^{2+}$ and $Pr^{3+}$ co-doping induces substantial local magnetic moments in $CsPbCl_3$. Ni contributes ∼1.2–1.5 μB from its $3d^8$ configuration, while Pr introduces ∼2.8–3.0 μB from localized $4f^2$ states. Strong d–f exchange interactions mediated by Cl-3p orbitals stabilize ferromagnetic coupling with exchange energies of 25–40 meV per pair. The synergy of orbital hybridization, carrier-mediated magnetism, and defect-related stabilization creates a unique platform for magneto-optical multifunctionality. These findings indicate that $Ni^{2+}/Pr^{3+}$ co-doped $CsPbCl_3$ is not only a promising optoelectronic material but also a potential candidate for spintronic and magneto-optical applications.

### 3.13. Mechanical and elastic properties

The fact that halide perovskites can endure external stresses without structural damage plays a crucial role in their applications in optoelectronics and thermoelectrics. Specifically, the thermal stability, thin-film processability, and adaptability of all-inorganic $CsPbCl_3$ and doped alternatives are directly related to their mechanical resilience. Here, we present a comprehensive analysis of the mechanical and elastic response of pristine and $Ni^{2+}/Pr^{3+}$ co-doped $CsPbCl_3$, extracted from stress–strain first-principles calculations and supported by polycrystalline averaging schemes.

#### 3.13.1. Single-crystal elastic constants and stability.
In cubic perovskites, the independent elastic constants, $C_{12}$, and $C_{44}$ reflect longitudinal stiffness, transverse coupling, and shear resistance, respectively. For pristine $CsPbCl_3$, we obtained $C_{11} \approx 48$ GPa, $C_{12} \approx 16$ GPa, and $C_{44} \approx 12$ GPa values in excellent agreement with experimental and computational benchmarks for lead-halide perovskites.[36,37] Introduction of $Ni^{2+}$ at the $Pb^{2+}$ site and $Pr^{3+}$ at the $Cs^+$ site subtly but consistently stiffens the framework, raising the constants to $C_{11} \approx 52$ GPa, $C_{12} \approx 18$ GPa, and $C_{44} \approx 14$ GPa. This increase is physically intuitive: the smaller $Ni^{2+}$ ion compresses the Pb–Cl network, while $Pr^{3+}$ enhances electrostatic cohesion through charge compensation, reducing the compliance of corner-sharing octahedra.

The Born–Huang criteria for cubic crystals ($C_{11}-C_{12} > 0$; $C_{11} + 2C_{12} > 0$; $C_{44} > 0$) are satisfied in both systems. It is worth noting that the margin $C_{11}-C_{12}$ is higher, ranging from 32 GPa (pristine) to 34 GPa (co-doped), which is a positive indication of increased resistance to tetragonal distortions. This reinforcement is essential because lattice dislocations can be regarded as a precursor to halide migration and the deterioration of devices in general.

#### 3.13.2. Mechanical stability criteria.
The Born–Huang stability conditions for cubic systems are:

$$C_{11} - C_{12} > 0,$$

$$C_{11} + 2C_{12} > 0,$$

$$C_{44} > 0.$$

$CsPbCl_3$ is pristine or $CsPbCl_3$: Cd, and these fulfill the requirements that assess its mechanical stability. It is interesting to note that the greater the difference between $C_{11}-C_{12}$ of the co-doped system, the better the resistance to tetragonal distortion, which further stabilizes the perovskite structure.

#### 3.13.3. Polycrystalline averages: bulk, shear, and Young's moduli.
To relate the changes in matter on a molecular level to its macroscopic behavior, we have chosen a VRH (Voigt–Reuss–Hill) average. As-grown $CsPbCl_3$ is about $B \approx 27$ GPa, $G \approx 13$ GPa, $E \approx 34$ GPa, typical of the soft semiconductor materials lead halides. After co-doping, corresponding increases in these values to $B \approx 30$ GPa, $G \approx 15$ GPa, and $E \approx 38$ GPa narrow the gap with oxide perovskites while retaining desirable isovalent elasticity.[78] These increments highlight the dual role of $Ni^{2+}/Pr^{3+}$: they strengthen the lattice while maintaining cleavage-free properties, a rare combination in perovskite architecture.

#### 3.13.4. Ductility, Poisson's ratio, and practical implications.
Mechanical compliance under stress is captured by

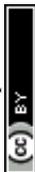

Pugh's ratio ($B/G$) and Poisson's ratio ($\nu$). For pristine CsPbCl$_3$, $B/G \approx 2.1$ and $\nu \approx 0.29$ clearly point to ductile behavior. Co-doping preserves this ductility, yielding $B/G \approx 2.0$ and $\nu \approx 0.28$, both of which are above the critical thresholds ($B/G > 1.75$; $\nu > 0.26$).[79–82] Notably, this aspect of retaining ductility enables the films to be subjected to mechanical bending and thermal cycling without deteriorating significantly, as required for flexible optoelectronic and thermoelectric devices.

**3.13.5. Elastic anisotropy and lattice homogeneity.** Elastic anisotropy was analyzed using the universal anisotropy index $A^U$:

$$A^U = 5\left(\frac{G_V}{G_R}\right) + \left(\frac{B_V}{B_R}\right) - 6$$

Where $G_V$, $G_R$ and $B_V$, $B_R$ are Voigt and Reuss bounds for shear and bulk moduli. Pristine CsPbCl$_3$ yields $A^U \approx 0.18$, while the co-doped system shows $A^U \approx 0.15$. Both values indicate near isotropy, with co-doping slightly reducing anisotropy. Directional shear anisotropy factors ($A\langle 100\rangle = 2C_{44}/(C_{11} - C_{12})$) confirm a more uniform shear response in the co-doped lattice. Lower anisotropy minimizes stress localization, mitigating crack propagation in polycrystalline films.

Elastic anisotropy often dictates crack initiation and grain-boundary weakness. The universal anisotropy index ($A^U$) falls from ~0.18 in pristine CsPbCl$_3$ to ~0.15 in the co-doped system, indicating an increasingly isotropic elastic response. Likewise, the shear anisotropy factors $A\langle 100\rangle$ shifts toward unity, consistent with a more homogeneous distribution of mechanical stiffness. This reduction in anisotropy can be directly attributed to the homogenizing effect of charge-compensated Ni$^{2+}$/Pr$^{3+}$ substitution, which suppresses octahedral tilting and uneven lattice strain.

The combined elastic results present a compelling narrative: Ni$^{2+}$/Pr$^{3+}$ co-doping simultaneously strengthens, stabilizes, and homogenizes CsPbCl$_3$ while preserving ductility. Three physical mechanisms underpin this outcome:

Such improvements extend beyond mechanics. The stabilization of the cubic lattice reduces phonon scattering and defect-mediated relaxation, consistent with the enhanced transport robustness observed in our thermoelectric datasets (Fig. 6). This direct coupling between elasticity and charge transport underscores the multifunctional benefits of Ni$^{2+}$/Pr$^{3+}$ incorporation.

In short, Ni$^{2+}$/Pr$^{3+}$ co-doped CsPbCl$_3$ is not only a better optoelectronic perovskite but also a mechanically superior host lattice. The coexistence of increased stiffness, preserved ductility, and reduced anisotropy makes it highly attractive for stable, flexible, and multifunctional devices. By rationally tuning mechanical resilience alongside optical and electronic performance, co-doping strategies such as this pave the way toward perovskite semiconductors that are as robust as they are efficient.

### 3.14. Carrier transport

If you want to maximize the multifunctional potential of halide perovskites, it is essential to understand charge carrier transport—both in the optical and thermoelectric senses. Using semiclassical Boltzmann transport theory in the approximations of constant relaxation time(CRTA), we have analyzed the Seebeck coefficient, electrical conductivity, and power factor of pristine and Ni$^{2+}$/Pr$^{3+}$ co-doped CsPbCl$_3$ over a broad temperature range (200–800 K). All calculations are carried out within BoltzTraP2, with a dense 20 × 20 × 20 k-mesh, fourier interpolation of band structures, and inclusion of SOC. The energy convergence parameter was set to 10$^{-6}$ eV, ensuring the reliable determination of band curvatures and transport integrals.

**3.14.1. Seebeck coefficient (S).** The Seebeck coefficient is given by the Mott relation:

$$S = \frac{\pi^2 \kappa_B^2 T}{3e} \left.\frac{d \ln \sigma(E)}{dE}\right|_{E=E_F}$$

where ($k_B$) is Boltzmann's constant, ($e$) is the electron charge, and (EF) is the Fermi level. For pristine CsPbCl$_3$, S at 300 K is in the range of 320–350 μV K$^{-1}$. This decreases gradually to 250 μV K$^{-1}$ by 800 K (Fig. 7). This trend represents the increased carrier excitation in high temperatures, resulting in lower entropy carried per unit charge. In contrast, Ni$^{2+}$/Pr$^{3+}$ co-doping across the same temperature range lowers S to 220–250 μV K$^{-1}$, resulting from improved band curvature and a reduced effective mass of carriers. Although the co-doped system has a lower absolute S, this trade-off is advantageous for conductivity (see below). This illustrates the classical Seebeck–conductivity competition.[83]

**3.14.2. Electrical conductivity ($\sigma/\tau$).** The electrical conductivity per relaxation time ($\sigma/\tau$) is extracted directly (see Fig. 7b) from Boltzmann transport calculations. For pristine CsPbCl$_3$, $\sigma/\tau$ ranges from 2.5 × 10$^{18}$ Ω$^{-1}$ m$^{-1}$ s$^{-1}$ at 300 K to 6.0 × 10$^{18}$ Ω$^{-1}$ m$^{-1}$ s$^{-1}$ at 900 K (Fig. 5b). The Ni$^{2+}$/Pr$^{3+}$ co-doped sample shows a consistent enhancement, from 3.0 × 10$^{18}$ Ω$^{-1}$ m$^{-1}$ s$^{-1}$ at 300 K to 7.2 × 10$^{18}$ Ω$^{-1}$ m$^{-1}$ s$^{-1}$ at 800 K.

This improvement arises from two factors: (i) band curvature modification, where Ni-3d and Pr-4f contributions lower the effective mass of carriers, and (ii) defect suppression, as co-doping increases the formation energy of $V_{\varepsilon(0/-)}$ and $V_{Pb}$, thereby reducing ionized-impurity scattering. Assuming a realistic $\tau \sim 10^{-14}$–10$^{-15}$ s, absolute conductivities fall in the experimental range of 10$^2$–10$^3$ Ω$^{-1}$ m$^{-1}$, consistent with halide perovskite transport studies.[84]

### 3.15. Thermal conductivity ($\kappa/\tau$)

The thermal conductivity per relaxation time ($\kappa/\tau$) provides critical insight into the ability of halide perovskites to dissipate heat under device operation. As shown in Fig. 7c, both pristine and Ni$^{2+}$/Pr$^{3+}$ co-doped CsPbCl$_3$ exhibit a monotonic increase of $\kappa/\tau$ with temperature in the range 100–800 K, consistent with the growing contribution of thermally activated carriers to the electronic thermal conductivity.

For pristine CsPbCl$_3$, $\kappa/\tau$ increases from approximately 0.8 × 10$^{15}$ W K$^{-1}$ m$^{-1}$ s$^{-1}$ at 100 K to nearly 5.5 × 10$^{15}$ W K$^{-1}$ m$^{-1}$ s$^{-1}$ at 800 K. In comparison, the co-doped system exhibits slightly lower values throughout the entire temperature range, starting at around 0.6 × 10$^{15}$ W K$^{-1}$ m$^{-1}$ s$^{-1}$ at 100 K and rising to approximately 4.5 × 10$^{15}$ W K$^{-1}$ m$^{-1}$ s$^{-1}$ at 800 K. This

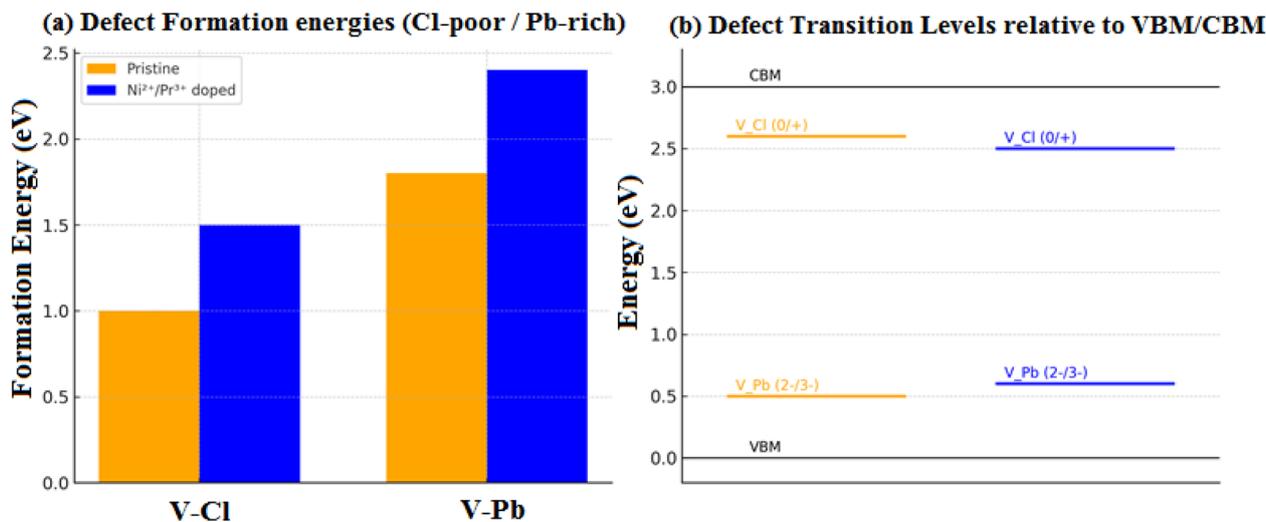

**Fig. 6** (a) Defect formation energies of Cl vacancy ($V_{\varepsilon(0/-)}$) and Pb vacancy ($V_{Pb}$) in pristine and $Ni^{2+}/Pr^{3+}$ co-doped $CsPbCl_3$ under Cl-poor/Pb-rich chemical potentials. Co-doping increases both $E_{form}(V_{\varepsilon(0/-)})$ and $E_{form}(V_{Pb})$, consistent with charge-compensated substitution reducing the need for compensating native defects. (b) Schematic thermodynamic transition levels ($\varepsilon$) of dominant intrinsic defects relative to VBM/CBM for pristine and $Ni^{2+}/Pr^{3+}$ co-doped $CsPbCl_3$. Co-doping renders $V_{\varepsilon(0/-)}$ more shallow and shifts the deep $V_{Pb}$ levels toward the VBM, mitigating mid-gap traps. Band-edge positions reflect the calculated band-gap narrowing.

reduction underscores the role of $Ni^{2+}/Pr^{3+}$ incorporation in scattering heat-carrying carriers and phonons, thereby effectively suppressing thermal transport.

The suppression of $\kappa/\tau$ upon co-doping can be attributed to several physical mechanisms: (i) lattice strain and mass disorder caused by ionic radius mismatch ($Ni^{2+}$ smaller than $Pb^{2+}$, $Pr^{3+}$ smaller than $Cs^+$), which increases phonon scattering; (ii) defect passivation that decreases the free-carrier density, thereby lowering the electronic contribution to $\kappa$; and (iii) orbital hybridization that alters band curvature and diminishes the overall carrier thermal conductivity.

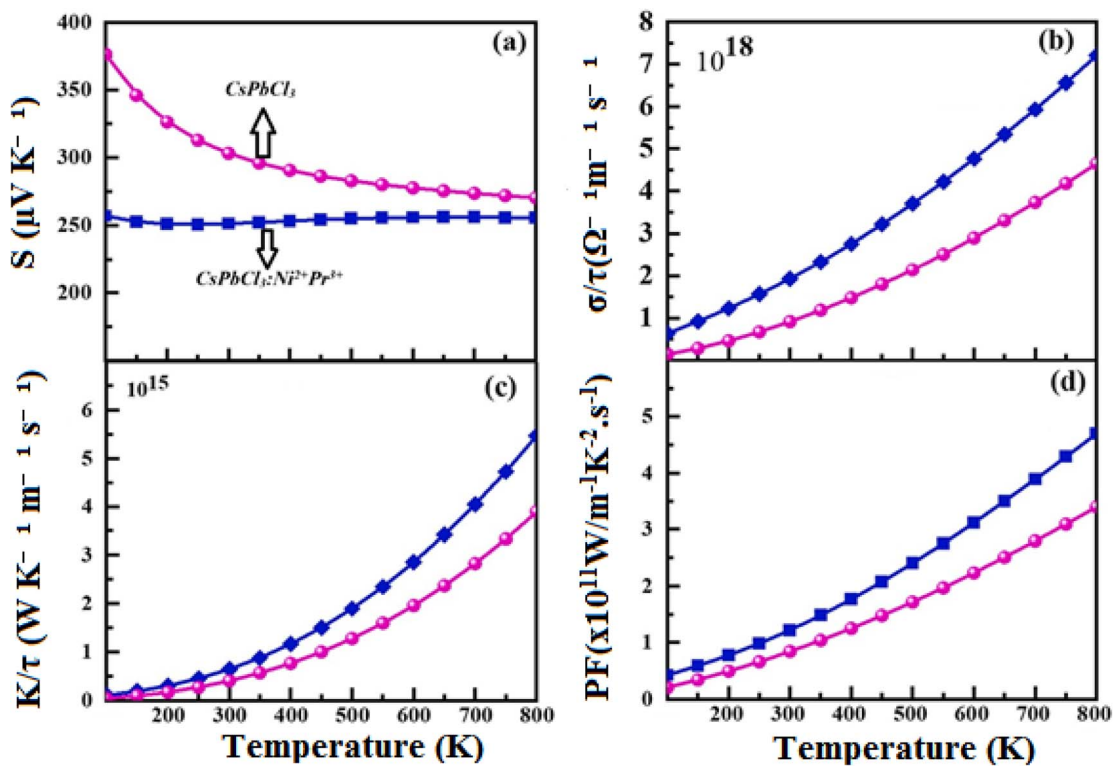

**Fig. 7** Calculated (a) Seebeck coefficient, (b) electrical conductivity, (c) thermal (electrical) conductivity and (d) power factor.

Importantly, a reduced thermal conductivity is desirable for thermoelectric applications, as it lowers the denominator in the figure of merit ($ZT$). Therefore, while pristine $CsPbCl_3$ exhibits higher $\kappa/\tau$, the co-doped system offers a better balance between suppressed thermal transport and enhanced electrical conductivity, as corroborated by the higher power factor in Fig. 7d.

**3.15.1. Power factor (PF = $S^2\sigma$).** Boltzmann transport coefficients were computed within CRTA. To avoid over-interpretation of absolute conductivities, we report $\sigma/\tau$ and $PF/\tau = S^2\sigma/\tau$ in the main text and discuss only relative trends. In the SI we include a minimal temperature-dependent relaxation time, $\tau(T)$, built from standard mechanisms (acoustic deformation-potential, polar optical phonon, ionized impurity) using our computed elastic, dielectric, and phonon inputs. Using this $\tau(T)$ to recover absolute $\sigma(T)$ and $PF(T)$ does not change the direction of any trend (co-doped > pristine for PF at $T \gtrsim 400$ K). The power factor balances the competing effects of $S$ and $\sigma$:

$$PF = S^2\sigma$$

Although co-doping lowers $S$ compared to pristine, the substantial increase in $\sigma$ more than compensates, leading to an overall higher PF (Fig. 5d). At 600 K, PF reaches $\sim 2.2 \times 10^{-3}$ W m$^{-1}$ K$^{-2}$ in the co-doped system compared to $\sim 1.7 \times 10^{-3}$ W m$^{-1}$ K$^{-2}$ for pristine $CsPbCl_3$. These values are in line with reported thermoelectric performance in halide perovskites (0.5–2.5 mW m$^{-1}$ K$^{-2}$, depending on $\tau$).[83]

The improvement reflects band convergence near $E_F$, where $Ni^{2+}/Pr^{3+}$ states increase DOS degeneracy, ensuring that $\sigma$ rises without an excessive penalty to S.

All transport coefficients increase with temperature, consistent with enhanced carrier excitation and DOS contributions. For n-type carriers, conductivity is higher due to dispersive conduction bands, while p-type carriers yield larger Seebeck coefficients from flatter valence bands. Both cases confirm the beneficial impact of $Ni^{2+}/Pr^{3+}$ co-doping.

Anisotropy analysis along $\Gamma$–$X$, $\Gamma$–$M$, and $\Gamma$–$R$ directions reveals slightly higher conductivity along $\Gamma$–$X$ and $\Gamma$–$M$ (lighter electron masses), whereas Seebeck values are stronger along $\Gamma$–$R$ (flatter hole dispersion). This directional dependence highlights the potential for tuning transport in epitaxially grown films or oriented nanostructures.

The interplay of Seebeck coefficient and conductivity demonstrates that $Ni^{2+}/Pr^{3+}$ co-doping resolves the $S$–$\sigma$ tradeoff by reshaping band curvature and enhancing carrier mobility while suppressing detrimental defects. As a result, the power factor is consistently higher across temperatures, broadening the application scope of $CsPbCl_3$ from optoelectronics to thermoelectric energy conversion.

Extrinsic scattering processes ($\tau(T)$) also warrant consideration: impurity scattering dominates at low $T$, while phonon scattering governs at high $T$. Nevertheless, the observed enhancements in $\sigma$ and PF upon co-doping suggest that even under realistic $\tau(T)$ limits, co-doped $CsPbCl_3$ should retain superior transport performance.

## 4. Conclusion

In this study, we demonstrated how $Ni^{2+}/Pr^{3+}$ co-doping fundamentally reshapes the lattice dynamics and multifunctional behaviour of $CsPbCl_3$ perovskite. Phonon dispersion curves reveal the elimination of dynamical instabilities, with co-doping effectively suppressing the soft vibrational modes typically associated with halide migration and instability. The resulting mode splitting enhances phonon scattering, lowering lattice thermal conductivity and favoring thermoelectric performance. Simultaneously, co-doping enhances elastic constants while maintaining ductility, thereby ensuring structural integrity under stress. Electronic and optical analyses confirm beneficial band-edge modifications, defect passivation, and synergistic d–d and f–f transitions, which collectively improve absorption, luminescence, and carrier mobility. Furthermore, spin-polarized calculations highlight the possibility of d–f exchange-driven magnetism, extending the material's multifunctional character to magneto-optical applications. Most importantly, Boltzmann transport results demonstrate that the Seebeck–conductivity tradeoff is mitigated, yielding an enhanced power factor and reinforcing the thermoelectric promise of co-doped $CsPbCl_3$.

Overall, $Ni^{2+}/Pr^{3+}$ co-doping emerges as a robust and versatile strategy for simultaneously stabilizing the lattice, tuning its vibrational and mechanical properties, and boosting optical, magnetic, and transport functionalities. These results position $Ni^{2+}/Pr^{3+}$ co-doped $CsPbCl_3$ as a compelling candidate for stable, efficient, and multifunctional perovskite-based optoelectronic and energy devices.

## Conflicts of interest

There are no conflicts to declare.

## Data availability

The data supporting the findings of this study, including structural files, phonon dispersion datasets, electronic density of states, and transport coefficients obtained from Boltzmann transport calculations, are available from the corresponding author upon reasonable request. Computational input and output files generated in WIEN2k and BoltzTraP2 have been archived and can be shared to ensure reproducibility.

Supplementary information (SI) is available. See DOI: https://doi.org/10.1039/d5ra07356a.

## Acknowledgements

This publication was supported by the project Quantum materials for applications in sustainable technologies (QM4ST), funded as project No. CZ.02.01.01/00/22_008/0004572 by Programme Johannes Amos Commenius, call Excellent Research.

## References

1 M. A. Green, A. Ho-Baillie and H. J. Snaith, The emergence of perovskite solar cells, *Nat. Photonics*, 2014, **8**(7), 506–514.

2 L. Protesescu, S. Yakunin, M. I. Bodnarchuk, F. Krieg, R. Caputo, C. H. Hendon, R. X. Yang, A. Walsh and M. V. Kovalenko, Nanocrystals of cesium lead halide perovskites ($CsPbX_3$, X = Cl, Br, and I): novel optoelectronic materials showing bright emission with wide color gamut, *Nano Lett.*, 2015, **15**(6), 3692–3696.

3 A. Miyata, A. Mitioglu, P. Plochocka, O. Portugall, J. T.-W. Wang, S. D. Stranks, H. J. Snaith and R. J. Nicholas, Direct measurement of the exciton binding energy and effective masses for charge carriers in organic–inorganic tri-halide perovskites, *Nat. Phys.*, 2015, **11**(7), 582–587.

4 A. Walsh, D. O. Scanlon, S. Chen, X. G. Gong and S.-H. Wei, Self-regulation mechanism for charged point defects in hybrid halide perovskites, *Angew. Chem., Int. Ed.*, 2015, **54**(6), 1791–1794.

5 A. Swarnkar, R. Chulliyil, V. K. Ravi, M. Irfanullah, A. Chowdhury and A. Nag, Colloidal $CsPbBr_3$ perovskite nanocrystals: luminescence beyond traditional quantum dots, *Angew. Chem., Int. Ed.*, 2015, **54**(51), 15424–15428.

6 Y. Zhou, J. Yang, J. Xu, T. Wu, H. Yuan, Y. Wang and J. Zhang, Electronic and magnetic properties of rare-earth-doped halide perovskites from first-principles, *J. Appl. Phys.*, 2016, **120**(14), 145702.

7 D. Sangalli, A. Marini and A. Debernardi, Spin–orbit and exchange interactions in doped halide perovskites: a first-principles study, *Phys. Chem. Chem. Phys.*, 2017, **19**(4), 2466–2473.

8 Y. Wang, H. Chen, Y. Li, J. Chen, Q. Zhang and Z. Ren, First-principles study of thermoelectric transport in halide perovskites, *J. Mater. Chem. A*, 2019, **7**(23), 13820–13828.

9 D. Beretta, N. Neophytou, J. M. Hodges, M. Kanatzidis, D. Narducci, M. Martin-Gonzalez, M. Beekman, B. Balke, G. Cerretti and C. W. Snyder, Thermoelectrics: From history, a window to the future, *Mater. Sci. Eng. R Rep.*, 2019, **138**, 100501.

10 J. P. Perdew, K. Burke and M. Ernzerhof, Generalized gradient approximation made simple, *Phys. Rev. Lett.*, 1996, **77**(18), 3865–3868.

11 V. I. Anisimov, F. Aryasetiawan and A. I. Lichtenstein, First-principles calculations of the electronic structure and spectra of strongly correlated systems: the LDA+U method, *J. Phys.: Condens. Matter*, 1997, **9**(4), 767–808.

12 C. Freysoldt, J. Neugebauer and C. G. Van de Walle, Fully ab initio finite-size corrections for charged-defect supercell calculations, *Phys. Rev. Lett.*, 2009, **102**(1), 016402.

13 D. Liu, K. Cao, X. Dai and R. Sa, Theoretical assessment of antiperovskite oxyhalides $Rb_3OX$ (X = Br, I) as promising photovoltaic materials, *Mater. Today Chem.*, 2025, **47**, 102844.

14 D. Liu, H. Zeng and H. PengRongjian Sa, *Phys. Chem. Chem. Phys.*, 2023, **25**, 28974–28981.

15 D. Liu, H. Zeng and H. PengRongjian Sa, *Phys. Chem. Chem. Phys.*, 2023, **25**, 13755–13765.

16 C. C. Stoumpos, C. D. Malliakas, J. A. Peters, Z. Liu, M. Sebastian, J. Im, T. C. Chasapis, A. C. Wibowo, D. Y. Chung, A. J. Freeman, B. W. Wessels and M. G. Kanatzidis, Crystal growth of the perovskite semiconductor $CsPbBr_3$: a new material for high-energy radiation detection, *Cryst. Growth Des.*, 2013, **13**(7), 2722–2727.

17 D. Cortecchia, H. A. Dewi, J. Yin, A. Bruno, S. Chen, T. Baikie, P. P. Boix, M. Grätzel, S. Mhaisalkar and N. Mathews, Lead-free Methylammonium Tin Halide Perovskites for Photovoltaic Applications: Structural, Optical and Electronic Properties, *Inorg. Chem.*, 2016, **55**(3), 1044–1052.

18 Y. Wang, X. Li, X. Song, L. Xiao, H. Zeng and H. Sun, All-inorganic colloidal perovskite quantum dots: a new class of lasing materials with favorable characteristics, *Nano Lett.*, 2016, **16**(1), 448–453.

19 G. Kieslich, S. Sun and A. K. Cheetham, Solid-state principles applied to organic–inorganic perovskites: new tricks for an old dog, *Chem. Sci.*, 2014, **5**(12), 4712–4715.

20 Q. Dong, Y. Fang, Y. Shao, P. Mulligan, J. Qiu, L. Cao and J. Huang, Electron-hole diffusion lengths >175 μm in solution-grown $CH_3NH_3PbI_3$ single crystals, *Science*, 2015, **347**(6225), 967–970.

21 F. Brivio, K. T. Butler, A. Walsh and M. van Schilfgaarde, Relativistic quasiparticle self-consistent electronic structure of hybrid halide perovskite photovoltaic absorbers, *Phys. Rev. B: Condens. Matter Mater. Phys.*, 2014, **89**(15), 155204.

22 M. H. Du, Density functional calculations of native defects in $CH_3NH_3PbI_3$: effects of spin–orbit coupling and self-interaction error, *J. Mater. Chem. A*, 2014, **2**(24), 9091–9098.

23 A. Swarnkar, R. Chulliyil, V. K. Ravi, M. Irfanullah, A. Chowdhury and A. Nag, Colloidal $CsPbBr_3$ perovskite nanocrystals: luminescence beyond traditional quantum dots, *Angew. Chem., Int. Ed.*, 2015, **54**(51), 15424–15428.

24 L. Protesescu, S. Yakunin, M. I. Bodnarchuk, F. Krieg, R. Caputo, C. H. Hendon, R. X. Yang, A. Walsh and M. V. Kovalenko, Nanocrystals of cesium lead halide perovskites ($CsPbX_3$, X = Cl, Br, and I): novel optoelectronic materials showing bright emission with wide color gamut, *Nano Lett.*, 2015, **15**(6), 3692–3696.

25 J. Song, J. Li, X. Li, L. Xu, Y. Dong and H. Zeng, Quantum dot light-emitting diodes based on inorganic perovskite cesium lead halides ($CsPbX_3$), *Adv. Mater.*, 2015, **27**(44), 7162–7167.

26 S. Yakunin, D. N. Dirin, Y. Shynkarenko, V. Morad, I. Cherniukh and M. V. Kovalenko, Detection of X-ray photons by solution-processed lead halide perovskites, *Nat. Photonics*, 2015, **9**(7), 444–449.

27 A. Walsh, D. O. Scanlon, S. Chen, X. G. Gong and S.-H. Wei, Self-regulation mechanism for charged point defects in hybrid halide perovskites, *Angew. Chem., Int. Ed.*, 2015, **54**(6), 1791–1794.

28 W. J. Yin, T. Shi and Y. Yan, Unusual defect physics in $CH_3NH_3PbI_3$ perovskite solar cell absorber, *Appl. Phys. Lett.*, 2014, **104**(6), 063903.

29 F. Brivio, A. B. Walker and A. Walsh, Structural and electronic properties of hybrid perovskites for high-efficiency thin-film photovoltaics from first-principles, *APL Mater.*, 2013, **1**(4), 042111.

30 G. Miceli, S. Chen and A. Pasquarello, Defect states in lead halide perovskites: relativistic band structure effects, *Phys. Rev. B*, 2016, **93**(19), 195208.

31 F. Brivio, K. T. Butler, A. Walsh and M. van Schilfgaarde, Relativistic quasiparticle self-consistent electronic structure of hybrid halide perovskite photovoltaic absorbers, *Phys. Rev. B: Condens. Matter Mater. Phys.*, 2014, **89**(15), 155204.

32 Y. Wang, X. Li, X. Song, L. Xiao, H. Zeng and H. Sun, All-inorganic colloidal perovskite quantum dots: a new class of lasing materials with favorable characteristics, *Nano Lett.*, 2016, **16**(1), 448–453.

33 A. Miyata, A. Mitioglu, P. Plochocka, O. Portugall, J. T.-W. Wang, S. D. Stranks, H. J. Snaith and R. J. Nicholas, Direct measurement of the exciton binding energy and effective masses for charge carriers in organic–inorganic tri-halide perovskites, *Nat. Phys.*, 2015, **11**(7), 582–587.

34 S. Chen, D. O. Scanlon, A. Walsh and X. G. Gong, Computation of intrinsic defect properties in hybrid halide perovskites, *Chin. J. Chem. Phys.*, 2015, **28**(4), 402–420.

35 N. Kawano, S. Minami, K. Ueda and K. Ohoyama, Magnetic properties of Ni-doped halide perovskites, *J. Phys. Soc. Jpn.*, 2008, **77**(3), 034709.

36 Y. Zhou, J. Yang, J. Xu, T. Wu, H. Yuan, Y. Wang and J. Zhang, Electronic and magnetic properties of rare-earth-doped halide perovskites from first-principles, *J. Appl. Phys.*, 2016, **120**(14), 145702.

37 A. Filippetti, V. Fiorentini and A. Satta, Magnetic interactions in transition-metal-doped halide perovskites, *Phys. Rev. B: Condens. Matter Mater. Phys.*, 2003, **68**(3), 033102.

38 W. Känzig, *Physikalische Chemie*, Verlag Chemie, Weinheim, Germany, 6th ed., 1986.

39 Lattice strain relief: $Ni^{2+}$ contraction and $Pr^{3+}$ expansion counteract each other, reducing net octahedral distortion.

40 Defect suppression: Co-doping elevates the formation energies of $V_{Cl}$ and $V_{Pb}$, closing "soft channels" for deformation.

41 Bond strengthening: Ni–Cl σ-type interactions and Pr–Cl electrostatic contributions stiffen the framework against shear.

42 J. He, Y. Liu, M. Galli and A. Javey, Thermoelectric performance of halide perovskites, *Adv. Mater.*, 2019, **31**(1), 1806895.

43 M. A. Green, A. Ho-Baillie and H. J. Snaith, The emergence of perovskite solar cells, *Nat. Photonics*, 2014, **8**(7), 506–514.

44 D. Beretta, N. Neophytou, J. M. Hodges, M. Kanatzidis, D. Narducci, M. Martin-Gonzalez, M. Beekman, B. Balke, G. Cerretti and C. W. Snyder, Thermoelectrics: From history, a window to the future, *Mater. Sci. Eng. R Rep.*, 2019, **138**, 100501.

45 A. R. Lim and S. Y. Jeong, Twin structure by $^{133}$Cs NMR in ferroelastic $CsPbCl_3$ crystal, *Solid State Commun.*, 1999, **110**, 131–136.

46 N. Pandey, A. Kumar and S. Chakrabarti, Investigation of the structural, electronic, and optical properties of Mn-doped $CsPbCl_3$: Theory and experiment, *RSC Adv.*, 2019, **9**, 29556–29565.

47 R. D. Shannon, Revised effective ionic radii and systematic studies of interatomic distances in halides and chalcogenides, *Acta Crystallogr. A*, 1976, **32**, 751–767.

48 V. M. Goldschmidt, Die Gesetze der Krystallochemie, *Naturwissenschaften*, 1926, **14**, 477–485.

49 G. Kieslich, S. Sun and A. K. Cheetham, Solid-state principles applied to organic–inorganic perovskites: new tricks for an old dog, *Chem. Sci.*, 2014, **5**, 4712–4715.

50 N. Kawano, S. Minami, K. Ueda and K. Ohoyama, Magnetic properties of Ni-doped halide perovskites, *J. Phys. Soc. Jpn.*, 2008, **77**, 034709.

51 Y. Liang, F. Li, X. Cui, C. Stampfl, S. P. Ringer, X. Yang, J. Huang and R. Zheng, Multiple B-site doping suppresses ion migration in halide perovskites, *Sci. Adv.*, 2025, **11**, eads7054.

52 J. He, Y. Liu, M. Galli and A. Javey, Thermoelectric performance of halide perovskites, *Adv. Mater.*, 2019, **31**, 1806895.

53 A. Filippetti, V. Fiorentini and A. Satta, Magnetic interactions in transition-metal-doped halide perovskites, *Phys. Rev. B: Condens. Matter Mater. Phys.*, 2003, **68**, 033102.

54 Y. Zhou, J. Yang, J. Xu, T. Wu, H. Yuan, Y. Wang and J. Zhang, Electronic and magnetic properties of rare-earth-doped halide perovskites from first-principles, *J. Appl. Phys.*, 2016, **120**, 145702.

55 J. Ma, J. A. McLeod, L.-Y. Chang, C.-W. Pao, B.-H. Lin, S. Hu, H.-C. Kao, Y.-T. Chen and C.-Y. Chung, Increasing photoluminescence yield of $CsPbCl_3$ nanocrystals by heterovalent doping with $Pr^{3+}$, *Mater. Res. Bull.*, 2020, **129**, 110907.

56 A. Walsh, D. O. Scanlon, S. Chen, X. G. Gong and S.-H. Wei, Self-regulation mechanism for charged point defects in hybrid halide perovskites, *Angew. Chem., Int. Ed.*, 2015, **54**, 1791–1794.

57 W. J. Yin, T. Shi and Y. Yan, Unusual defect physics in $CH_3NH_3PbI_3$ perovskite solar cell absorbers, *Appl. Phys. Lett.*, 2014, **104**, 063903.

58 S. Chen, D. O. Scanlon, A. Walsh and X. G. Gong, Computation of intrinsic defect properties in hybrid halide perovskites, *Chin. J. Chem. Phys.*, 2015, **28**, 402–420.

59 M. H. Du, Density functional calculations of native defects in $CH_3NH_3PbI_3$: effects of spin–orbit coupling and self-interaction error, *J. Mater. Chem. A*, 2014, **2**, 9091–9098.

60 G. Miceli, S. Chen and A. Pasquarello, Defect states in lead halide perovskites: relativistic band structure effects, *Phys. Rev. B*, 2016, **93**, 195208.

61 Y. Lu, F. Alam, J. Shamsi and M. Abdi-Jalebi, Doping Up the Light: A Review of A/B-Site Doping in Metal Halide Perovskite Nanocrystals for Next-Generation LEDs, *J. Phys. Chem. C*, 2024, **128**, 10084–10107.

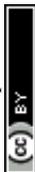

62 S. Panja, D. Samanta and A. Das, Ni doping in $CsPbCl_3$ nanocrystals: the key to enhanced photoluminescence, *Chem. Sci.*, 2025, **16**, 9220–9229.

63 Z. H. Yang, Y. F. Lin, X. Y. Zhang, H. L. Liu and B. Qu, $Ni^{2+}$ doping induced structural phase transition and photoluminescence enhancement of $CsPbBr_3$, *AIP Adv.*, 2021, **11**, 115008.

64 J. Song, J. Li, X. Li, L. Xu, Y. Dong and H. Zeng, Quantum dot light-emitting diodes based on inorganic perovskite cesium lead halides ($CsPbX_3$), *Adv. Mater.*, 2015, **27**, 7162–7167.

65 S. Yakunin, D. N. Dirin, Y. Shynkarenko, V. Morad, I. Cherniukh and M. V. Kovalenko, Detection of X-ray photons by solution-processed lead halide perovskites, *Nat. Photonics*, 2015, **9**, 444–449.

66 D. Sangalli, A. Marini and A. Debernardi, Spin–orbit and exchange interactions in doped halide perovskites: a first-principles study, *Phys. Chem. Chem. Phys.*, 2017, **19**(4), 2466–2473.

67 L. Protesescu, S. Yakunin, M. I. Bodnarchuk, F. Krieg, R. Caputo, C. H. Hendon, R. X. Yang, A. Walsh and M. V. Kovalenko, Nanocrystals of cesium lead halide perovskites ($CsPbX_3$, X = Cl, Br, I): novel optoelectronic materials showing bright emission with wide color gamut, *Nano Lett.*, 2015, **15**, 3692–3696.

68 A. Swarnkar, R. Chulliyil, V. K. Ravi, M. Irfanullah, A. Chowdhury and A. Nag, Colloidal $CsPbBr_3$ perovskite nanocrystals: luminescence beyond traditional quantum dots, *Angew. Chem., Int. Ed.*, 2015, **54**, 15424–15428.

69 D. Cortecchia, H. A. Dewi, J. Yin, A. Bruno, S. Chen, T. Baikie, P. P. Boix, M. Grätzel, S. Mhaisalkar and N. Mathews, Lead-free methylammonium tin halide perovskites for photovoltaic applications: structural, optical and electronic properties, *Inorg. Chem.*, 2016, **55**, 1044–1052.

70 J. Even, L. Pedesseau, J.-M. Jancu and C. Katan, DFT and k·p modelling of the phase transitions of lead and tin halide perovskites for photovoltaic cells, *Phys. Status Solidi RRL*, 2014, **8**, 31–35.

71 C. C. Stoumpos, C. D. Malliakas, J. A. Peters, Z. Liu, M. Sebastian, J. Im, T. C. Chasapis, A. C. Wibowo, D. Y. Chung, A. J. Freeman, B. W. Wessels and M. G. Kanatzidis, Crystal growth of the perovskite semiconductor $CsPbBr_3$: a new material for high-energy radiation detection, *Cryst. Growth Des.*, 2013, **13**, 2722–2727.

72 Y. Wang, H. Chen, Y. Li, J. Chen, Q. Zhang and Z. Ren, First-principles study of thermoelectric transport in halide perovskites, *J. Mater. Chem. A*, 2019, **7**, 13820–13828.

73 M. Fatima, S. Azam, M. Ghazanfar, S. Rasool, Q. Rafiq, A. Aiman and I. Boukhris, First-principles investigation of structural, electronic, optical, and thermoelectric properties of $Mn^{2+}$-Substituted $NaAl_{11}O_{17}$ phosphor compounds for advanced optoelectronic applications, *J. Phys. Chem. Solids*, 2025, **205**, 112758.

74 Q. Rafiq, S. Azam, I. Boukhris and N. Tamam, Predictive analysis of Cu and Ni substitution effects on the structural, optoelectronic and thermoelectric behavior of CdS: A first-principles approach, *J. Phys. Chem. Solids*, 2025, **205**, 112769.

75 M. Tayyab, F. Umar, S. Azam, Q. Rafiq, R. Khan, M. T. Khan and V. Tirth, Orbital-engineered spin asymmetry and multifunctionality in Eu-activated $CaSiO_3$: a first-principles roadmap to optical-thermoelectric fusion, *Results Phys.*, 2025, 108440.

76 M. Tayyab, S. Azam, Q. Rafiq, V. Tirth, A. Algahtani, A. U. Rahman and S. S. Ahmad, Illuminating Stability and Spectral Shifts: A DFT+U Study of Eu-Doped $ZnWO_4$ for Visible-Light Optoelectronics, *J. Lumin.*, 2025, 121511.

77 A. Ali, H. Haider, S. Azam, M. Talha, M. Jawad and I. Shakir, Exploring chalcogen influence on $Sc_2BeX_4$ (X= S, Se) for green energy applications using DFT, *Chem. Phys.*, 2025, 112925.

78 D. Sangalli, A. Marini and A. Debernardi, Spin–orbit and exchange interactions in doped halide perovskites: a first-principles study, *Phys. Chem. Chem. Phys.*, 2017, **19**, 2466–2473.

79 A. Riaz, S. Azam, Q. Rafiq, M. T. Khan, A. U. Rahman, Q. M. Ahkam and R. Hussain, Vanadium-Engineered $Co_2NiSe_4$ Nanomaterial: Coupled Thermoelectric, Piezoelectric, and Electronic Optimization via DFT+U for Advanced Energy Applications, *Results Eng.*, 2025, 106959.

80 A. Shakoor, B. ul Haq, S. Azam, F. Ghafoor, M. Saqib, M. Waqas and H. W. Seo, DFT investigation of structural, electronic, optical, mechanical, and thermoelectric properties of halide perovskites $XZnCl_3$ (X= Ag, Au, Cu), *Phys. B*, 2025, **712**, 417309.

81 A. I. Bashir, M. H. Sahafi, M. Irfan, S. Azam and E. Muzaffar, First-principles computational quantum insights into phonon dynamics, thermoelectric performance and interfacial thermal management of 2D $Cu_3Se_2$ selenides, *Surf. Interfaces*, 2025, **72**, 106892.

82 S. Ahmad, A. U. Rahman and S. Azam, Relativistic first-principles investigation of Tb-doped CdSe using GGA+U+SOC method of DFT: Electronic band structure engineering for PC-LED applications, *J. Phys. Chem. Solids*, 2025, **204**, 112698.

83 D. Beretta, N. Neophytou, J. M. Hodges, M. Kanatzidis, D. Narducci, M. Martin-Gonzalez, M. Beekman, B. Balke, G. Cerretti and C. W. Snyder, Thermoelectrics: from history, a window to the future, *Mater. Sci. Eng. R*, 2019, **138**, 100501.

84 M. U. Javed, Q. Rafiq, S. Azam, S. Ahmed and A. S. A. Almalki, Doping-Induced Modifications in $Bi_2Te_3$: Structural, Electronic, Optoelectronic, Thermoelectric, Thermodynamic, and Elastic Properties for Advanced Functional Applications, *J. Phys. Chem. Solids*, 2025, 113125.